\documentclass[%
 aip,
 jmp,%
 amsmath,amssymb,
preprint,%
 reprint,%
author-year,%
author-numerical,%
]{revtex4-2}

\usepackage{graphicx}% Include figure files
\usepackage{dcolumn}% Align table columns on decimal point
\usepackage{bm}% bold math

\begin{document}

%------------------------------------------------

%\def\CTeXPreproc{Created by ctex v0.2.14, don't edit!}\documentclass [12pt,twoside] {article}
%\documentclass [twoside,12pt] {article}
%\usepackage{setspace,graphicx,bm,fancyhdr}
%\usepackage{amsmath}
%\usepackage{amsmath,amsthm,amssymb,latexsym}
%\usepackage{color}
%\usepackage{graphicx}

%\usepackage{authblk}
%\renewcommand{\baselinestretch}{1.2}
%\setlength{\topmargin}{-0.2in}
%\setlength{\textwidth}{6in}
%\setlength{\textheight}{8.5in}
%\setlength{\oddsidemargin}{0.25in}
%\setlength{\evensidemargin}{0.25in}
%\newcommand{\mbf}[1]{\mathbf{#1}}
%\newcommand{\bl}{\color{blue}}
%\newcommand{\mbv}[1]{\mbox{\boldmath$#1$\unboldmath}}
%\newcommand{\rf}{\vskip .1in\par\sloppy\hangindent=1pc\hangafter=1
%                 \noindent}
\newcommand{\chp}[1]{\mbox{$\stackrel{\wedge}{#1}$}}
\newcommand{\ichp}[1]{\mbox{$\stackrel{\vee}{#1}$}}
\newcommand{\jchp}[1]{\mbox{$\stackrel{\_}{#1}$}}
\newcommand{\ch}[1]{\mbox{$\stackrel{\sim}{#1}$}}
\newcommand{\vct}[1]{\mbox{$\stackrel{\rightarrow}{#1}$}}
\newcommand{\ft}[1]{\mbox{$\stackrel{\wedge}{#1}$}}
\newcommand{\slas}[1]{\mbox{${{#1} \!\!\! /}$}}
\newcommand{\Mob}[0]{M\"{o}bius }

%\raggedbottom

%\parskip 0.05in
%\begin{document}

\preprint{AIP/123-QED}

\title[Vertex function]{UV and IR divergence-free calculation of the vertex function at arbitrary values of its arguments}

\author{John Mashford}
 \altaffiliation[ ]{School of Mathematics and Statistics \\
University of Melbourne, Victoria 3010, Australia \\
E-mail: mashford@unimelb.edu.au}%Lines break automatically or can be forced with \\
%\email{Second.Author@institution.edu.}
%\affiliation{ 
%Authors' institution and/or address%\\This line break forced with \textbackslash\textbackslash
%}

 \homepage{https://findanexpert.unimelb.edu.au/profile/11242-john-mashford}
%\affiliation{%
%Second institution and/or address%\\This line break forced% with \\
%}

\date{\today}% It is always \today, today,

\begin{abstract}
The vertex function is analyzed using covariant spectral regularization without encountering any divergence, either UV or IR. The mathematics of covariant  spectral regularization for covariant matrix valued measures with one Lorentz index on open subsets of Minkowski space is described. This is then applied to the case of the vertex function and expressions for the densities associated with the vertex function in the t channel and the s channel with respect to Lebesgue measure on Minkowski space are obtained. These densities are well defined, non-divergent and analytic over their domains of definition and are obtained without using renormalization or needing to consider final state radiation. The limit of the expression for the vertex function in the t channel at low energy and low momenta is computed resulting in the classical result for the leading order (LO) contribution to the anomalous magnetic moment of the electron. Also the density for the vertex function in the s channel is used to compute the  LO vertex correction contribution to the high energy limit of the cross section for the process $e^{+}e^{-}\rightarrow\mu^{+}\mu^{-}$. 
\end{abstract}

%Use showkeys class option if keyword
\maketitle

%\title{\bf } 
%\author{} \\
%}
%\date{\today}
%\maketitle
%\newcommand{\rf}{\vskip .1in\par\sloppy\hangindent=1pc\hangafter=1 \noindent}
%\newcommand{\bl} 
%\raggedbottom

\tableofcontents

\section{Introduction}

The calculation of the vertex function is a computation of great importance in quantum field theory (QFT) and has played an important role in QFT since its inception. The principal problem with the Feynman integral associated with the vertex function and other related Feynman integrals is its IR divergence \cite{Jauch,Piazza}.

IR divergence is a significant problem in many areas of physics, from Yang-Mills theory \cite{Ding,Butler}, cosmology \cite{Tanaka} and quantum gravity \cite{Wilson-Gerow} to high energy physics \cite{Yennie}.

The Feynman integral associated with the vertex function is both UV and IR divergent when treated in the conventional manner, i.e. when viewed as defining a  function pointwise. The UV divergence is typically removed using Pauli-Villars or dimensional regularization together with the method of renormalization. The IR divergence can be avoided by introducing a photon mass $m_{\gamma}$ for the virtual photon, however this parameter cannot be eliminated by renormalization. For example, in the high energy limit, the cross section involves the Sudakov double logarithm which remains after differences between cross sections at different scales are computed.

As well as occurring for some loop diagram Feynman integrals, IR divergence is associated with certain tree level processes involving initial or final state radiation (e.g. bremsstrahlung). An example is the process $e^{-}e^{-}\rightarrow e^{-}e^{-}\gamma$, i.e. M\o ller scattering with final state radiation. The IR divergences for such processes can be removed by giving the final state photon a fictitious mass $m_{\gamma}$. 
It has been found that when one computes the vertex correction contribution to the cross sections for such processes, the terms involving a virtual photon mass $m_{\gamma}$ precisely cancel the terms involving a fictitious final state photon mass $m_{\gamma}$. This leads to the method currently used for removing IR divergences in QFT. One must include soft photon final state (or initial state) radiation in order to obtain finite cross sections. 

It is usually stated \cite{Schwartz,Weinberg} that the method described above for removing IR divergences is physically justified since cross sections associated with, for example, $e^{-}e^{-}\rightarrow e^{-}e^{-}$ are not physically observable {\em per se}. Processes with different final states must be included. It is physically impossible to determine whether a final state is just an electron or an electron plus a number of soft or collinear photons. No experiment can observe soft photons at arbitrarily low energies, any detector has a finite resolution.

The above described approach is formalized in the Bloch-Nordsieck theorem \cite{Bloch} which says that, given a finite energy resolution, IR divergences will always cancel in QED if finite state radiation is included. In QCD this theorem is not true and needs to be modified \cite{Doria,Catani}. One has, in general for unitary theories, the KLN theorem \cite{Kinoshita,Lee,Akhoury,Khalil,Frye} which says that, for such theories, IR divergences will cancel when all possible final and initial states are summed with respect to a finite energy window. 
This constitutes the general approach currently used to deal with IR divergences in QFT \cite{Nakanishi,Zhou,Chung,Cvitanovic}.

We have developed a method of regularization which we have called spectral regularization (see Refs. \cite{Springer,Symmetry,IJMPA,NPB}) in which problematic objects in QFT are viewed as covariant complex vector, matrix or tensor valued measures on Minkowski space or subsets of Minkowski space, their spectra are computed using a spectral calculus and the densities associated with these spectra are used in QFT calculations. We now call this technique covariant spectral regularization in order to distinguish it from other techniques called ``spectral regularization".

With covariant spectral regularization renormalization is not required to cancel UV divergences and final state radiation does not need to be considered in order to cancel IR divergences, since there are no divergences, UV or IR.

In the present paper we apply covariant spectral regularization to the vertex function and compute analytic matrix valued densities (with one Lorentz index) defined for arbitrary arguments for this object in the t channel and the s channel without using renormalization or adding in computations associated with final state radiation.

The resulting densities which we have obtained without encountering either UV or IR divergence, can be used in QFT calculations. As an example, we compute from the t channel vertex function the leading order  (LO) contribution to the anomalous magnetic moment of the electron. We also compute from the s channel vertex function the LO contribution to the high energy limit of the vertex correction contribution to the cross section for the process $e^{+}e^{-}\rightarrow\mu^{+}\mu^{-}$ without needing to include final state radiation to cancel IR divergence.

In Section~\ref{section:maths} we present some of the mathematical techniques and formalism involved in covariant spectral regularization. In Section~\ref{section:vertex_function} we apply these techniques to the case of the vertex function in the t channel and compute the density associated with this object. In Section~\ref{section:anom_mag_mom} we use the results of Section~\ref{section:vertex_function} to give a simple derivation of the expression for the LO  contribution to the anomalous magnetic moment of the electron. In Section \ref{section:vertex_function_s} we compute the density associated with the vertex function in the s channel and in Section \ref{section:electron_positron_annihilation} we use this density in the calculation of the LO contribution to the high energy limit of the vertex correction contribution to the cross section for the process $e^{+}e^{-}\rightarrow\mu^{+}\mu^{-}$. Our computation of the cross section for this process results in precisely the same prediction for the total cross section as the standard prediction. However our differential cross section differs in detail from the standard prediction. In Section~\ref{section:differential_cross_section_comparison} we discuss the differences between these two predictions. The paper concludes with Section~\ref{section:conclusion}. 

\section{$K$ covariant ${\bf C}^{4\times4}$ valued measures with one Lorentz index\label{section:maths}}

Let $K\subset U(2,2)$ be the group
\begin{equation}
K=\left\{\left(\begin{array}{cc}
a&0\\
0&a^{\dagger-1}
\end{array}\right):a\in GL(2,{\bf C}),|\mbox{det}(a)|=1\right\}.
\end{equation}
$K$ is locally isomorphic to $SL(2,{\bf C})\times U(1)$ and, as well as acting on ${\bf C}^4$ in the usual way, it acts on ${\bf R}^4$ in a natural way according to  \cite{AMP}
\begin{align*}
\kappa p&=\Lambda(\kappa)p\mbox{ for }p\in{\bf R}^4,
\end{align*}
where $\Lambda(\kappa)$ is the Lorentz transformation corresponding to $\kappa\in K$. $\kappa\mapsto\Lambda(\kappa)$ is a homomorphism from $K$ to the proper orthochronous Lorentz group $O(1,3)^{\uparrow+}$.

Let $U$ be an open subset of Minkowski space such that $O(1,3)^{\uparrow+}U=U$, i.e. $\Lambda p\in U,\forall\Lambda\in O(1,3)^{\uparrow+},p\in U$.

Let ${\mathcal B}_0(U)=\{\Upsilon\in{\mathcal B}(U):\overline{\Upsilon}\subset U\mbox{ and $\overline{\Upsilon}$ is compact}\}$ where ${\mathcal B}(U)$ denotes the Borel algebra of $U$ and $\overline\Upsilon$ denotes the closure of $\Upsilon$. By a Borel complex measure on $U$ we will mean a map $\mu:{\mathcal B}_0(U)\rightarrow{\bf C}$ such that for all $C\in{\mathcal B}_0(U)$ the map $\mu|_C:{\mathcal B}(C)\rightarrow{\bf C}$ defined by $\mu|_C(\Upsilon)=\mu(\Upsilon)$ is a Borel complex measure on $C$ in the usual sense \cite{Halmos} (and is hence finite on $C$). For the rest of this paper the term `measure' will mean `Borel measure'. 

A complex matrix valued measure with one index $\Phi^{\mu}:{\mathcal B}_0(U)\rightarrow{\bf C}^{4\times4}$ will be said to be $K$ covariant if
\begin{equation} \label{eq:condition1}
\Phi^{\mu}(\kappa\Upsilon)={\Lambda^{\mu}}_{\nu}\kappa\Phi^{\nu}(\Upsilon)\kappa^{-1}, \forall\kappa\in K,\Upsilon\in{\mathcal B}_0(U),
\end{equation}
where $\Lambda=\Lambda(\kappa)$ is the Lorentz transformation corresponding to $\kappa\in K$ \cite{AMP,IJMPA}.

\subsection{Existence of spectral function when measure is absolutely continuous with respect to Lebesgue measure}

Suppose that $\Phi^{\mu}:{\mathcal B}_0(U)\rightarrow{\bf C}^{4\times4}$ is a $K$ covariant complex matrix valued measure with one Lorentz index which can be generated by a locally integrable density which we will, without fear of confusion, denote by $\Phi^{\mu}$ also. Then
\begin{equation} \label{eq:density_measure}
\Phi^{\mu}(\Upsilon)=\int_{\Upsilon}\Phi^{\mu}(p)\,dp, \forall\Upsilon\in{\mathcal B}_0(U),
\end{equation}
and
\begin{eqnarray}
\Phi^{\mu}(\kappa\Upsilon) & = & \int_{\Lambda(\Upsilon)}\Phi^{\mu}(p)\,dp=\int_{\Upsilon}\Phi^{\mu}(\Lambda p)\,dp,
\end{eqnarray}
for all $\kappa\in K,\Lambda=\Lambda(\kappa)\in O(1,3)^{\uparrow+},\Upsilon\in{\mathcal B}_0(U)$ where we have used the Lorentz invariance of the Lebesgue measure. Therefore by Eq.~\eqref{eq:condition1}
\begin{equation}
\int_{\Upsilon}\Phi^{\mu}(\Lambda p)\,dp={\Lambda^{\mu}}_{\nu}\kappa\int_{\Upsilon}\Phi^{\nu}(p)\,dp\,\kappa^{-1}=\int_{\Upsilon}{\Lambda^{\mu}}_{\nu}\kappa\Phi^{\nu}(p)\kappa^{-1}\,dp,
\end{equation}
for all $\kappa\in K,\Upsilon\in{\mathcal B}_0(U)$. Since this is true for all $\Upsilon\in{\mathcal B}_0(U)$ we must have that for all $\kappa\in K$
\begin{equation} \label{eq:condition2}
\Phi^{\mu}(\Lambda p)={\Lambda^{\mu}}_{\nu}\kappa\Phi^{\nu}(p)\kappa^{-1},
\end{equation}
for almost all $p\in U$. We will consider the (non-pathological) case where $\Phi^{\mu}$ can be (and has been) adusted on a set of measure zero so that Eq.~\eqref{eq:condition2} holds for all $\kappa\in K,p\in U$.  

Conversely if a matrix valued locally integrable function with one index $\Phi^{\mu}:U\rightarrow{\bf C}^{4\times4}$ satisfies Eq.~\eqref{eq:condition2} then the object $\Phi^{\mu}:{\mathcal B}_0(U)\rightarrow{\bf C}^{4\times4}$ defined by Eq.~\eqref{eq:density_measure} is a $K$ covariant complex matrix valued measure with one Lorentz index.

Thus we are interested in locally integrable functions $\Phi^{\mu}:U\rightarrow{\bf C}^{4\times4}$ which satisfy
\begin{equation} \label{eq:condition3}
\Phi^{\mu}(\kappa p)=\Lambda^{\mu}{}_{\nu}\kappa\Phi^{\nu}(p)\kappa^{-1}, \forall\kappa\in K,\Lambda=\Lambda(\kappa)\in O(1,3)^{\uparrow+},p\in U,
\end{equation}
and we will call such functions $K$ covariant.
Clearly such a function is determined by its values on the set $\{(m,{\vct 0})\}:m\in{\bf R}\}\cup\{(0,m,0,0):m\in(0,\infty)\}\cup\{(1,1,0,0),(-1,1,0,0)\}$ (i.e. on representatives of the orbits of $O(1,3)^{\uparrow+}$ on Minkowski space). 

We will call a measure or a function causal if it is supported in $\{p\in{\bf R}^4:p^2\geq0,p^0\geq0\}$. 
If $\Phi^{\mu}$ is a causal $K$ covariant matrix valued function with one Lorentz index, define the function $M^{\mu}=M_{\Phi}^{\mu}:(0,\infty)\rightarrow{\bf C}^{4\times4},\mu=0,1,2,3,$ by
\begin{equation}
M^{\mu}(m)=\Phi^{\mu}((m,{\vct0})),
\end{equation}
where we set $M^{\mu}(m)=0$ if $(m,{\vct0}){\slas\in}U$.
We will call $M^{\mu}$ the spectrum of $\Phi^{\mu}$. $\Phi^{\mu}$ is, up to its values on the future null cone of the origin which is a set of Lebesgue measure zero, determined by its spectrum.

\subsection{Canonical form of causal $K$ covariant ${\bf C}^{4\times4}$ valued measures with one Lorentz index}

We embed ${\bf C}$ in ${\bf C}^4$ in the usual way, as scalar matrices.
Let $\sigma_i:{\mathcal B}_0([0,\infty))\rightarrow{\bf C}$ for $i=1,\ldots,4$ be complex measures. Define $\Phi^{\mu}:{\mathcal B}_0({\bf R}^4)\rightarrow{\bf C}^{4\times4}$ by
\begin{align}
\Phi^{\mu}(\Upsilon)=&\int_{m=0}^{\infty}\int_{{\bf R}^4}\chi_{\Upsilon}(p)\gamma^{\mu}\,\Omega_m^{+}(dp)\,\sigma_1(dm)\label{eq:canonical_form}\\
+&\int_{m=0}^{\infty}\int_{{\bf R}^4}\chi_{\Upsilon}(p){\slas p}\gamma^{\mu}\,\Omega_m^{+}(dp)\,\sigma_2(dm)\nonumber\\
+&\int_{m=0}^{\infty}\int_{{\bf R}^4}\chi_{\Upsilon}(p)p^{\mu}\,\Omega_m^{+}(dp)\,\sigma_3(dm)\nonumber\\
+&\int_{m=0}^{\infty}\int_{{\bf R}^4}\chi_{\Upsilon}(p){\slas p}p^{\mu}\,\Omega_m^{+}(dp)\,\sigma_4(dm),\nonumber\\
\end{align}
where $\chi_{\Upsilon}$ denotes the characteristic function of a set $\Upsilon$ defined by
\begin{equation}
\chi_{\Upsilon}(p)=\left\{\begin{array}{l}
1\mbox{ if }p\in\Upsilon\\
0\mbox{ otherwise,}
\end{array}\right.
\end{equation}
and we denote, for $m\geq0$, $\Omega^{\pm}_m$ to be the standard Lorentz invariant measure on the mass shell \cite{Symmetry,IJMPA} 
\begin{equation}
H^{+}_m=\{p\in{\bf R}^4:p^2=m^2,p^0\geq0\}\mbox{ or }H^{-}_m=\{p\in{\bf R}^4:p^2=m^2,p^0\leq0\}\mbox{ respectively},
\end{equation}
which satisfies
\begin{equation}\label{eq:Omega_def}
\int\psi(p)\,\Omega_m^{\pm}(dp)=\int\psi((\pm\omega_m({\vct p}),{\vct p}))\,\frac{d{\vct p}}{\omega_m({\vct p})},
\end{equation}
for all measurable functions $\psi:{\bf R}^4\rightarrow{\bf C}$ for which the integral on the right hand side of Eq.~\eqref{eq:Omega_def} exists.
Here, for $m\in{\bf R}$, $\omega_m:{\bf R}^3\rightarrow[0,\infty)$ is the function
\begin{equation}
\omega_m({\vct p})=(m^2+{\vct p}^2)^{\frac{1}{2}}.
\end{equation}

It is straightforword to show that $\Phi^{\mu}$ is a (Borel) complex matrix valued measure.

We will show that $\Phi^{\mu}$ defined by the first term only of Eq.~\eqref{eq:canonical_form} is a $K$ covariant matrix valued measure with one Lorentz index as follows.
Let 
\begin{equation}
\Phi_1^{\mu}(\Upsilon)=\int_{m=0}^{\infty}\int_{{\bf R}^4}\chi_{\Upsilon}(p)\gamma^{\mu}\,\Omega_m^{+}(dp)\,\sigma_1(dm)\mbox{ for }\Upsilon\in{\mathcal B}_0({\bf R}^4).
\end{equation}
Then
\begin{align*}
\Phi_1^{\mu}(\kappa\Upsilon)=&\int_{m=0}^{\infty}\int_{{\bf R}^4}\chi_{\kappa\Upsilon}(p)\gamma^{\mu}\,\Omega_m^{+}(dp)\,\sigma_1(dm)\\
=&\int_{m=0}^{\infty}\int_{{\bf R}^4}\chi_{\Upsilon}(\kappa^{-1}p)\gamma^{\mu}\,\Omega_m^{+}(dp)\,\sigma_1(dm)\\
=&\kappa\eta^{\mu\nu}\int_{m=0}^{\infty}\int_{{\bf R}^4}\chi_{\Upsilon}(p)\kappa^{-1}\gamma_{\nu}\kappa\,\Omega_m^{+}(dp)\,\sigma_1(dm)\,\kappa^{-1},
\end{align*}
where we have used the Lorentz invariance of $\Omega_m^{+}$. Now from the fundamental intertwining property of the Feynman slash $\Sigma(p)={\slas p}$ which is that \cite{AMP,IJMPA}
\begin{equation}\label{eq:intertwining_property}
\Sigma(\kappa p)=\kappa{\slas p}\kappa^{-1},\forall\kappa\in K,p\in{\bf R}^4,
\end{equation}
it follows that \cite{AMP,IJMPA} for all $\kappa\in K$
\[ \kappa^{-1}\gamma_{\nu}\kappa=\Lambda^{-1\rho}{}_{\nu}\gamma_{\rho}, \]
where $\Lambda=\Lambda(\kappa)$ is the Lorentz transformation corresponding to $\kappa$.
Also, since 
\[ \Lambda^{T}\eta\Lambda=\eta,\mbox{ which implies that }\eta\Lambda^{T}=\Lambda^{-1}\eta, \]
we have
\begin{equation}\label{eq:Lorentz_identity}
\eta^{\mu\nu}\Lambda^{-1\rho}{}_{\nu}=(\Lambda^{-1}\eta)^{\rho\mu}=(\eta\Lambda^{T})^{\rho\mu}=\eta^{\rho\nu}\Lambda^{T}{}_{\nu}{}^{\mu}=\eta^{\rho\nu}\Lambda^{\mu}{}_{\nu}.
\end{equation}
Thus
\begin{align*}
\Phi_1^{\mu}(\kappa\Upsilon)&=\kappa\eta^{\rho\nu}\Lambda^{\mu}{}_{\nu}\int_{m=0}^{\infty}\int_{{\bf R}^4}\chi_{\Upsilon}(p)\gamma_{\rho}\,\Omega_m^{+}(dp)\,\sigma_1(dm)\,\kappa^{-1}\\
&=\kappa\Lambda^{\mu}{}_{\nu}\int_{m=0}^{\infty}\int_{{\bf R}^4}\chi_{\Upsilon}(p)\gamma^{\nu}\,\Omega_m^{+}(dp)\,\sigma_1(dm)\,\kappa^{-1}\\
&=\Lambda^{\mu}{}_{\nu}\kappa\Phi_1^{\nu}(\Upsilon)\kappa^{-1},
\end{align*}
and so $\Phi_1^{\mu}$ is a $K$ covariant ${\bf C}^{4\times4}$ valued measure on Minkowski space with one Lorentz index as required.

We will now show that $\Phi^{\mu}$ defined by the second term only of Eq.~\eqref{eq:canonical_form} is a $K$ covariant matrix valued measure on Minkowski space with one Lorentz index as follows.
Let 
\[ \Phi_2^{\mu}(\Upsilon)=\int_{m=0}^{\infty}\int_{{\bf R}^4}\chi_{\Upsilon}(p){\slas p}\gamma^{\mu}\,\Omega_m^{+}(dp)\,\sigma_2(dm)\mbox{ for }\Upsilon\in{\mathcal B}_0({\bf R}^4). \]
Then
\begin{align*}
\Phi_2^{\mu}(\kappa\Upsilon)&=\int_{m=0}^{\infty}\int_{{\bf R}^4}\chi_{\kappa\Upsilon}(p){\slas p}\gamma^{\mu}\,\Omega_m^{+}(dp)\,\sigma_2(dm)\\
&=\int_{m=0}^{\infty}\int_{{\bf R}^4}\chi_{\Upsilon}(\kappa^{-1}p){\slas p}\gamma^{\mu}\,\Omega_m^{+}(dp)\,\sigma_2(dm)\\
&=\int_{m=0}^{\infty}\int_{{\bf R}^4}\chi_{\Upsilon}(p)\kappa{\slas p}\kappa^{-1}\gamma^{\mu}\,\Omega_m^{+}(dp)\,\sigma_2(dm)\\
&=\eta^{\mu\nu}\kappa\int_{m=0}^{\infty}\int_{{\bf R}^4}\chi_{\Upsilon}(p){\slas p}\kappa^{-1}\gamma_{\nu}\kappa\,\Omega_m^{+}(dp)\,\sigma_2(dm)\,\kappa^{-1}\\
&=\eta^{\mu\nu}\kappa\int_{m=0}^{\infty}\int_{{\bf R}^4}\chi_{\Upsilon}(p){\slas p}\Lambda^{-1\rho}{}_{\nu}\gamma_{\rho}\,\Omega_m^{+}(dp)\,\sigma_2(dm)\,\kappa^{-1}\\
&=\eta^{\rho\nu}\Lambda^{\mu}{}_{\nu}\kappa\int_{m=0}^{\infty}\int_{{\bf R}^4}\chi_{\Upsilon}(p){\slas p}\gamma_{\rho}\,\Omega_m^{+}(dp)\,\sigma_2(dm)\,\kappa^{-1}\\
&=\Lambda^{\mu}{}_{\nu}\kappa\int_{m=0}^{\infty}\int_{{\bf R}^4}\chi_{\Upsilon}(p){\slas p}\gamma^{\nu}\,\Omega_m^{+}(dp)\,\sigma_2(dm)\,\kappa^{-1}\\
&=\Lambda^{\mu}{}_{\nu}\kappa\Phi_2^{\nu}(\Upsilon)\kappa^{-1},
\end{align*}
where we have used the Lorentz invariance of $\Omega_m^{+}$, the fundamental intertwining property of the Feynman slash and Eq.~\eqref{eq:Lorentz_identity}. This proves the required result.

Similarly, the third and fourth terms of the canonical form given by Eq.~\eqref{eq:canonical_form} can be shown to be $K$ covariant ${\bf C}^{4\times4}$ valued measures with one Lorentz index. Therefore, given any complex measures $\sigma_1,\ldots,\sigma_4$ on $[0,\infty)$ a measure $\Phi^{\mu}$ of the form defined by Eq.~\eqref{eq:canonical_form} is a causal $K$ covariant ${\bf C}^{4\times4}$ vaued measure with one Lorentz index.

In particular, if $\sigma_1,\ldots,\sigma_4$ are locally integrable functions on $(0,\infty)$ then $\Phi^{\mu}$ is such a measure and is given by
\begin{align} \label{eq:canonical}
\Phi^{\mu}(\Upsilon)=&\int_{m=0}^{\infty}\int_{{\bf R}^4}\chi_{\Upsilon}(p)(\sigma_1(m)\gamma^{\mu}+\sigma_2(m){\slas p}\gamma^{\mu}+\nonumber\\
&\sigma_3(m)p^{\mu}+\sigma_4(m){\slas p}p^{\mu})\,\Omega_m(dp)\,dm,
\end{align}
where, for $m>0$, $\Omega_m=\Omega_m^{+}$.

Note that a measure of the form
\[ \Phi^{\mu}(\Upsilon)=\int_{m=0}^{\infty}\int_{{\bf R}^4}\chi_{\Upsilon}(p)\gamma^{\mu}{\slas p}\,\Omega_m^{+}(dp)\,\sigma(dm), \]
is $K$ covariant but, since
\[ \gamma^{\mu}{\slas p}=p_{\alpha}\gamma^{\mu}\gamma^{\alpha}=p_{\alpha}(2\eta^{\mu\alpha}-\gamma^{\alpha}\gamma^{\mu})=2p^{\mu}-{\slas p}\gamma^{\mu}, \]
such a measure is of the canonical form of Eq.~\eqref{eq:canonical_form}. Also, a measure of the form
\[ \Phi^{\mu}(\Upsilon)=\int_{m=0}^{\infty}\int_{{\bf R}^4}\chi_{\Upsilon}(p){\slas p}\gamma^{\mu}{\slas p}\,\Omega_m^{+}(dp)\,\sigma(dm), \]
is $K$ covariant but, since
\begin{align*}
{\slas p}\gamma^{\mu}{\slas p}=&p_{\alpha}p_{\beta}\gamma^{\alpha}\gamma^{\mu}\gamma^{\beta}\\
=&p_{\alpha}p_{\beta}\gamma^{\alpha}(2\eta^{\mu\beta}-\gamma^{\beta}\gamma^{\mu})\\
=&2{\slas p}p^{\mu}-{\slas p}{\slas p}\gamma^{\mu}\\
=&2{\slas p}p^{\mu}-p^2\gamma^{\mu},
\end{align*}
such a measure is of the form of the canonical form of Eq.~\eqref{eq:canonical_form}.

\subsection{Determination of the density for such measures when in canonical form}

Suppose that $\Phi^{\mu}$ is a measure of the form of Eq.~\eqref{eq:canonical}. Then
\begin{align*}
\Phi^{\mu}(\Upsilon)=&\int_{m=0}^{\infty}\int_{{\bf R}^3}\chi_{\Upsilon}((\omega_m({\vct p}),{\vct p}))(\sigma_1(m)\gamma^{\mu}+\sigma_2(m){\slas p}\gamma^{\mu}+\\
&\sigma_3(m)p^{\mu}+\sigma_4(m){\slas p}p^{\mu})\frac{d{\vct p}}{\omega_m({\vct p})}\,dm,
\end{align*}
where
$p=(\omega_m({\vct p}),{\vct p})$. 
Now make the coordinate transformation
\begin{equation}
q=q(m,{\vct p})=(\omega_m({\vct p}),{\vct p}).
\end{equation}
The Jacobian for the transformation is
\begin{equation}
J(m,{\vct p})=m\omega_m({\vct p})^{-1}.
\end{equation}
Therefore, since $m=(q^2)^{\frac{1}{2}}$, we have
\begin{align*}
\Phi^{\mu}(\Upsilon)=&\int_{q^2>0,q^0>0}\chi_{\Upsilon}(q)(\sigma_1(\zeta(q))\gamma^{\mu}+\sigma_2(\zeta(q)){\slas q}\gamma^{\mu}+\\
&\sigma_3(\zeta(q))q^{\mu}+\sigma_4(\zeta(q)){\slas q}q^{\mu})\frac{1}{\zeta(q)}\,dq,
\end{align*}
where $\zeta(q)=(q^2)^{\frac{1}{2}}$.
Therefore $\Phi^{\mu}$ is absolutely continuous with respect to Lebesgue measure on $\{q\in{\bf R}^4:q^2>0,q^0>0\}$ and the density corresponding to $\Phi^{\mu}$ is
\begin{align} \label{eq:density}
\Phi^{\mu}(q)=&Q^{-1}(\sigma_1(Q)\gamma^{\mu}+\sigma_2(Q){\slas q}\gamma^{\mu}+\nonumber\\
&\sigma_3(Q)q^{\mu}+\sigma_4(Q){\slas q}q^{\mu}),
\end{align}
where $Q=(q^2)^{\frac{1}{2}}$. 

The spectrum for such a density is given by
\begin{align}
M^{\mu}(Q)&=\Phi^{\mu}((Q,{\vct0}))\nonumber\\
&=Q^{-1}(\sigma_1(Q)\gamma^{\mu}+\sigma_2(Q)(Q\gamma^0)\gamma^{\mu}+\nonumber\\
&\sigma_3(Q)(Q\eta^{\mu0})+\sigma_4(Q)(Q\gamma^0)(Q\eta^{\mu0}))\nonumber\\
&=Q^{-1}\sigma_1(Q)\gamma^{\mu}+\sigma_2(Q)\gamma^0\gamma^{\mu}+\nonumber\\
&\sigma_3(Q)\eta^{\mu0}+Q\sigma_4(Q)\gamma^0\eta^{\mu0}\label{eq:spectrum},
\end{align} 
for $Q>0$.

\subsection{The spectral calculus for causal $K$ covariant ${\bf C}^{4\times4}$ valued measures with one Lorentz index}

Let $\Phi^{\mu}$ be a causal $K$ covariant measure with one Lorentz index and suppose that $\Phi^{\mu}$ is absolutely continuous with respect to Lebesgue measure on $\{p\in{\bf R}^4:p^2>0,p^0>0\}$ with a continuous density. Define, for $a,b,\epsilon>0,a<b,\mu=0,1,2,3$ the function 
\begin{equation}
g^{\mu}(a,b,\epsilon)=\Phi^{\mu}(\Upsilon(a,b,\epsilon)),
\end{equation}
where $\Upsilon(a,b,\epsilon)$ is the hyperbolic cylinder of radius $\epsilon$ between $a$ and $b$ defined by \cite{Springer,Symmetry,NPB}
\begin{equation}
\Upsilon(a,b,\epsilon)=\bigcup_{m\in(a,b)}S(m,\epsilon),
\end{equation}
where
\begin{equation}
S(m,\epsilon)=\{p\in{\bf R}^4:p^2=m^2,p^0>0,|{\vct p}|<\epsilon\}.
\end{equation}
Then
\begin{align*}
g^{\mu}(a,b,\epsilon)&=\int\chi_{\Upsilon(a,b,\epsilon)}(p)\Phi^{\mu}(p)\,dp\\
&\approx\int_{p^0=-\infty}^{\infty}\int_{{\vct p}\in{\bf R}^3}\chi_{(a,b)}(p^0)\chi_{B_{\epsilon}({\vct0})}({\vct p})\Phi^{\mu}((p^0,{\vct p}))\,d{\vct p}\,dp^0\\
&\approx\frac{4}{3}\pi\epsilon^3\int_{p^0=a}^b\Phi^{\mu}((p^0,{\vct0}))\,dp^0.
\end{align*}
Therefore, defining 
\[ g^{\mu}_a(b)=\lim_{\epsilon\rightarrow0}\epsilon^{-3}g^{\mu}(a,b,\epsilon), \]
we can compute the spectrum $M^{\mu}$ for $\Phi^{\mu}$ using the formula
\begin{equation}
M^{\mu}(b)=\frac{3}{4\pi}g_a^{\mu\prime}(b),b>a.
\end{equation}
This computation is exact in the limit as $\epsilon\rightarrow0$ \cite{Springer,Symmetry}.

\section{The vertex function in the t channel \label{section:vertex_function}}

\begin{figure} 
\centering
\includegraphics[width=6cm]{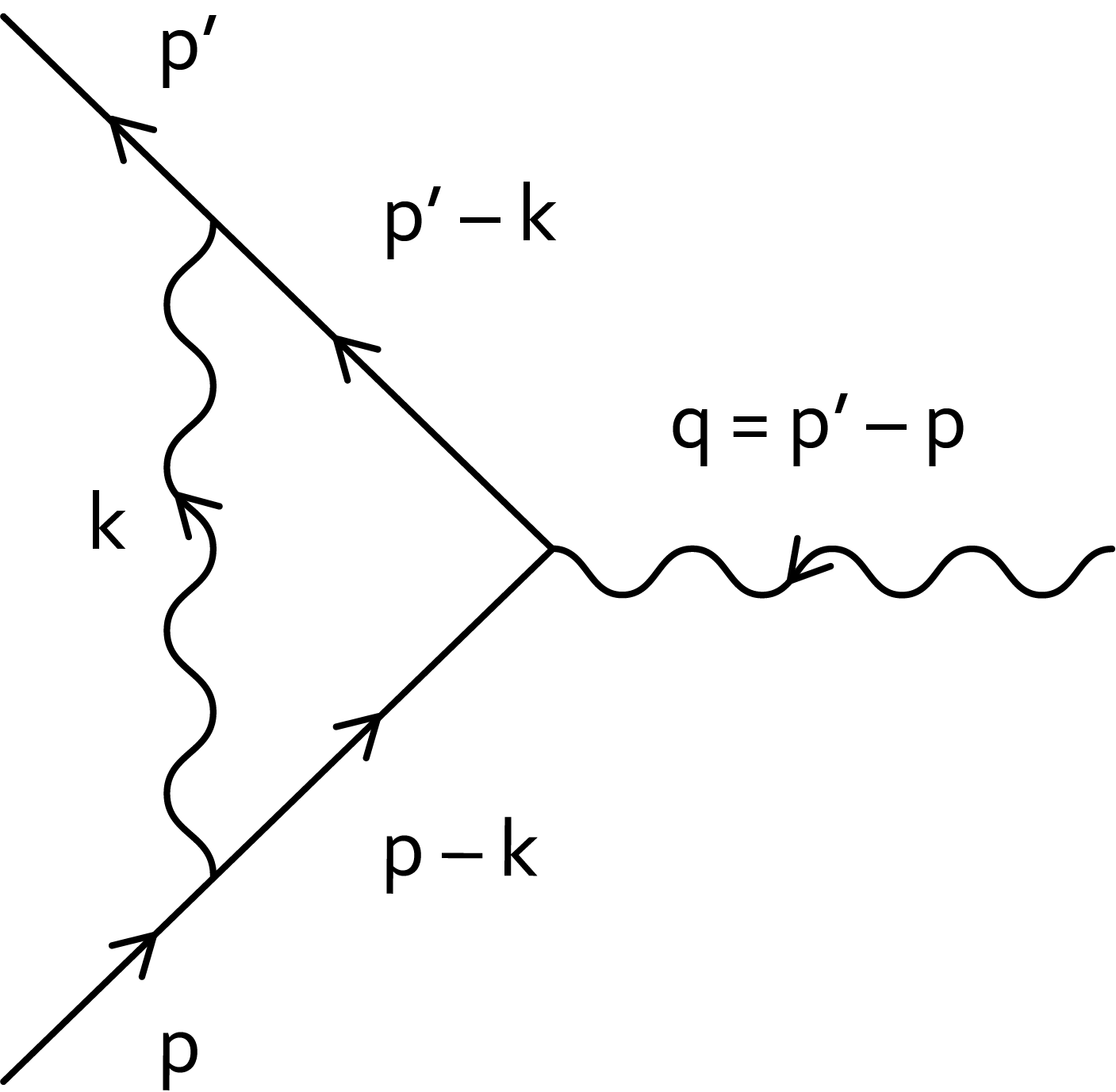}
\caption{Feynman diagram for the t channel vertex functionin QFT} \label{figure:QFT_vertex}
\end{figure}

Consider the Feynman diagram for the vertex function for QFT in the t channel shown in Figure~\ref{figure:QFT_vertex}. 
Applying the Feynman rules the vertex function is described by the Feynman integral
\begin{equation} \label{eq:vertex_modification_integral}
\Gamma^{\mu}=\int\frac{d^4k}{(2\pi)^4}iD_{\rho\sigma}(k)ie\gamma^{\rho}iS(p^{\prime}-k)ie\gamma^{\mu}iS(p-k)ie\gamma^{\sigma}.
\end{equation}
where
\begin{equation}
D_{\rho\sigma}(k)=\frac{-\eta_{\rho\sigma}}{k^2+i\epsilon},
\end{equation}
is the photon propagator,
\begin{equation}
S(p)=\frac{1}{{\slas p}-m+i\epsilon},
\end{equation}
is the fermion propagator, $e$ is the magnitude of the charge of the electron (or other fermion) and $m$ is the mass of the electron (or other fermion). 

\subsection{The t channel vertex function as a causal $K$ covariant ${\bf C}^{4\times4}$ valued measure with one Lorentz index}

From Eq.~\eqref{eq:vertex_modification_integral}
\begin{equation}
\Gamma^{\mu}(p^{\prime},p)=\frac{e^3}{(2\pi)^4}\int\frac{1}{k^2+i\epsilon}\gamma^{\nu}\frac{1}{{\slas p^{\prime}}-{\slas k}-m+i\epsilon}\gamma^{\mu}\frac{1}{{\slas p}-{\slas k}-m+i\epsilon}\gamma_{\nu}\,dk.
\end{equation}
The diagram of Figure \ref{figure:QFT_vertex} represents the vertex function associated with the scattering of, for example, an electron of mass $m$ from a proton. Therefore $p,p^{\prime}\in H_m$ where, for $m>0, H_m=H_m^{+}$. 
Hence there exist, by a well known theorem, a ${\vct p}\in{\bf R}^3$ such that
\begin{align*}
p=&(E,-{\vct p}),\\
p^{\prime}=&(E,{\vct p}),
\end{align*}
where $E=\omega_m({\vct p})$.

Now make a ``flip" transformation $F:({\bf R}^4)^2\rightarrow({\bf R}^4)^2$ defined by
\begin{equation}
F(p^{\prime},p)=(p^{\prime},Tp),
\end{equation}
where  $T:{\bf R}^4\rightarrow{\bf R}^4$  is the energy (time) inversion operator defined by
\begin{equation}
Tp=T(p^0,{\vct p})=(-p^0,{\vct p}),\forall p\in{\bf R}^4.
\end{equation}
Then
\begin{align*}
p=&(-E,-{\vct p}),\\
p^{\prime}=&(E,{\vct p}).
\end{align*}
Thus $p^{\prime}=\frac{1}{2}q,p=-\frac{1}{2}q$ where $q=p^{\prime}-p$. 

Therefore
\begin{equation}\label{eq:Phi_def}
\Gamma^{\mu}(p^{\prime},p)=\Phi^{\mu}(q),
\end{equation}
where
\begin{equation} \label{eq:Phi_def_t_channel}
\Phi^{\mu}(q)=\frac{e^3}{(2\pi)^4}\Xi^{\mu}(\frac{1}{2}q),
\end{equation}
with $\Xi^{\mu}$ given by
\begin{equation}\label{eq:Xi_def}
\Xi^{\mu}(q)=\int\frac{1}{k^2+i\epsilon}\gamma^{\nu}\frac{1}{{\slas q}-{\slas k}-m+i\epsilon}\gamma^{\mu}\frac{1}{-{\slas q}-{\slas k}-m+i\epsilon}\gamma_{\nu}\,dk.
\end{equation} 

The integral defining $\Xi^{\mu}(q)$ does not converge for any $q\in{\bf R}^4$ so regularization is required. We will use covariant spectral regularization. 

We will show, by a formal argument, that $\Xi^{\mu}$ can be interpreted as a $K$ covariant matrix valued measure with one Lorentz index as follows. 
Suppose (``pretend") that $\Xi^{\mu}$ existed as a ${\bf C}^{4\times4}$ valued function on Minkowski space. Then it is associated with a measure, which we shall also denote as $\Xi^{\mu}$, defined by
\begin{align*}
\Xi^{\mu}(\Upsilon)&=\int_{\Upsilon}\Xi^{\mu}(q)\,dq\\
&=\int\chi_{\Upsilon}(q)\Xi^{\mu}(q)\,dq\\
&=\int\chi_{\Upsilon}(q)(\int\frac{1}{k^2+i\epsilon}\gamma^{\nu}\frac{1}{{\slas q}-{\slas k}-m+i\epsilon}\gamma^{\mu}\frac{1}{-{\slas q}-{\slas k}-m+i\epsilon}\gamma_{\nu}\,dk)\,dq\\
&=\int\chi_{\Upsilon}(q)\frac{1}{k^2+i\epsilon}\gamma^{\nu}\frac{1}{{\slas q}-{\slas k}-m+i\epsilon}\gamma^{\mu}\frac{1}{-{\slas q}-{\slas k}-m+i\epsilon}\gamma_{\nu}\,dk\,dq\\
&"="\int\chi_{\Upsilon}(q)\frac{1}{k^2+i\epsilon}\gamma^{\nu}\frac{1}{{\slas q}-{\slas k}-m+i\epsilon}\gamma^{\mu}\frac{1}{-{\slas q}-{\slas k}-m+i\epsilon}\gamma_{\nu}\,dq\,dk\\
&=-\int\chi_{\Upsilon}(q+k)\frac{1}{k^2+i\epsilon}\gamma^{\nu}\frac{1}{{\slas q}-m+i\epsilon}\gamma^{\mu}\frac{1}{{\slas q}+2{\slas k}+m-i\epsilon}\gamma_{\nu}\,dq\,dk\\
&=-\int\chi_{\Upsilon}(q+k)\frac{1}{k^2+i\epsilon}\gamma^{\nu}\frac{{\slas q}+m}{q^2-m^2+i\epsilon}\gamma^{\mu}\frac{1}{{\slas q}+2{\slas k}+m}\gamma_{\nu}\,dq\,dk,
 \end{align*}
where we have enclosed the = sign in quotes at the only unjustified formal step (interchange of the order of integration). 

Note that there must be one or more unjustified formal steps in the argument because the ``function" defined by Eq.~\eqref{eq:Xi_def} is not well defined and one cannot go from something not well defined to something well defined by a rigorous argument consisting of a sequence of equalities.

Using the result \cite{Symmetry,IJMPA}
\begin{equation}\label{eq:ansatz}
\frac{1}{p^2-m^2\pm i\epsilon}\rightarrow-i\pi\Omega_m^{\pm},
\end{equation}
and using the standard argument for covariant spectral regularization \cite{Springer,Symmetry,IJMPA,NPB}, we write
\begin{equation}
\Xi^{\mu}(\Upsilon)=\pi^2\int\chi_{\Upsilon}(q+k)\gamma^{\nu}({\slas q}+m)\gamma^{\mu}\frac{1}{{\slas q}+2{\slas k}+m}\gamma_{\nu}\,\Omega_m(dq)\,\Omega_0^{+}(dk).
\end{equation}
Now, since $\Omega_m$ is supported on $H_m$ and $\Omega_0^{+}$ is supported on $H_0^{+}$ we may assume that $q^2=m^2$ and $k^2=0$. Then we have
\begin{align*}
\frac{1}{{\slas q}+2{\slas k}+m}&=({\slas q}+2{\slas k}+m)^{-1}=({\slas q}+2{\slas k}-m)({\slas q}+2{\slas k}-m)^{-1}({\slas q}+2{\slas k}+m)^{-1}\\
&=({\slas q}+2{\slas k}-m)[({\slas q}+2{\slas k}+m)({\slas q}+2{\slas k}-m)]^{-1}.
\end{align*}
Now
\begin{align*}
({\slas q}+2{\slas k}+m)({\slas q}+2{\slas k}-m)&=({\slas q}+2{\slas k})^2-m^2\\
&={\slas q}^2+4{\slas k}^2+2{\slas q}{\slas k}+2{\slas k}{\slas q}-m^2\\
&=2{\slas q}{\slas k}+2{\slas k}{\slas q},
\end{align*}
But
\[ {\slas q}{\slas k}+{\slas k}{\slas q}=q_{\alpha}k_{\beta}(\gamma^{\alpha}\gamma^{\beta}+\gamma^{\beta}\gamma^{\alpha})=2\eta^{\alpha\beta}q_{\alpha}k_{\beta}=2q.k. \]
Hence
\begin{equation}
\frac{1}{{\slas q}+2{\slas k}+m}=\frac{1}{4}({\slas q}+2{\slas k}-m)(q.k)^{-1}.
\end{equation}
Therefore
\begin{equation}\label{eq:Xi_mu_def}
\Xi^{\mu}(\Upsilon)=\frac{\pi^2}{4}\int\chi_{\Upsilon}(q+k)\gamma^{\nu}({\slas q}+m)\gamma^{\mu}({\slas q}+2{\slas k}-m)\gamma_{\nu}(q.k)^{-1}\,\Omega_m(dq)\,\Omega_0^{+}(dk).
\end{equation}
We may compute that, for all $k$ and $q$ for which $k^2=0$, $q^2=m^2$ and ${\vct k}\neq{\vct0}$, 
\[ q.k=\omega_m({\vct q})|{\vct k}|-{\vct q}.{\vct k}\geq\omega_m({\vct q})|{\vct k}|-|{\vct q}||{\vct k}|>|{\vct q}||{\vct k}|-|{\vct q}||{\vct k}|=0. \]
Therefore $(q.k)^{-1}$ does not ``blow up" except when $||{\vct k}||\rightarrow0$, where $||\mbox{ , }||$ denotes the Euclidean norm for ${\bf R}^4$. 

Let
\begin{equation}
C_m=\{p\in{\bf R}^4:p^2\geq m^2,p^0>0\}.
\end{equation}
Suppose that $\Upsilon$ is a compact subset of the interior $C_m^{o}=\{p\in{\bf R}^4:p^2>m^2,p^0>0\}$ of $C_m$. We will show that the integral defined by Eq.~\eqref{eq:Xi_mu_def} exists and is non-divergent. Since $\Upsilon$ is compact $\exists M>0$ such that $p^0<M,\forall p\in\Upsilon$. Let $S_1=\{q\in H_m:q^0\leq M\}$ and let $S_2=\{k\in H_0^{+}:k^0\leq M\}$. Then, if $q\in H_m$ and $k\in H_0^{+}$, then $q{\slas\in}S_1$ or $k{\slas\in}S_2\Rightarrow q+k{\slas\in}\Upsilon$. Since $\Upsilon\cap S_1=\emptyset$ and the positive function $g(p,q)=||p-q||$ is continuous on the compact set $\Upsilon\times S_1$ there exists an $a>0$ such that $||p-q||>a,\forall p\in\Upsilon,q\in S_1$. Let $S_3=\{k\in{\bf R}^4:||k||<a\}$.  Then $q\in S_1,k\in S_3\Rightarrow q+k{\slas\in}\Upsilon$. The continuous non-negative function $f(q,k)=|(q.k)|^{-1}$ achieves a maximum value $C\in[0,\infty)$ say on the compact  set $S_1\times(S_2\backslash S_3)$. Thus
\begin{align*}
&\int\chi_{\Upsilon}(q+k)||\gamma^{\nu}({\slas q}+m)\gamma^{\mu}({\slas q}+2{\slas k}-m)\gamma_{\nu}(q.k)^{-1}||\,\Omega_m(dq)\,\Omega_0^{+}(dk)\\
=&\int_{k\in S_2\backslash S_3}\int_{q\in S_1}\chi_{\Upsilon}(q+k)||\gamma^{\nu}({\slas q}+m)\gamma^{\mu}({\slas q}+2{\slas k}-m)\gamma_{\nu}||\,|(q.k)^{-1}|\,\Omega_m(dq)\,\Omega_0^{+}(dk)\\
\leq&\,C\int_{k\in S_2\backslash S_3}\int_{q\in S_1}\chi_{\Upsilon}(q+k)||\gamma^{\nu}({\slas q}+m)\gamma^{\mu}({\slas q}+2{\slas k}-m)\gamma_{\nu}||\,\Omega_m(dq)\,\Omega_0^{+}(dk)\\
\leq&\,C\int_{{\vct k}\in\pi(S_2\backslash S_3)}\int_{{\vct q}\in\pi(S_1)}||\gamma^{\nu}({\slas q}+m)\gamma^{\mu}({\slas q}+2{\slas k}-m)||\frac{d{\vct q}}{m}\frac{d{\vct k}}{a},\\
&\mbox{where }q=(\omega_m({\vct q}),{\vct q}),k=(|{\vct k}|,{\vct k})\\
<&\,\infty,
\end{align*}
where $||\mbox{ . }||$ is the usual matrix norm and $\pi:{\bf R}^4\rightarrow{\bf R}^3$ is the standard projection defined by $\pi(p)=\pi(p^0,{\vct p})={\vct p}$. The integral in the last line of the above computation is finite because it is the integral with respect to Lebesgue measure of a continuous function over a compact set.

Hence the integral given by Eq.~\eqref{eq:Xi_mu_def} defining $\Xi^{\mu}(\Upsilon)$ exists for any $\Upsilon\in{\mathcal B}_0(C_m^{o})$ and $\mu\in\{0,1,2,3\}$. It is straighforward to show that, for any $C\in{\mathcal B}_0(C_m^{o})$, the map $\Xi_C^{\mu}:{\mathcal B}_0(C)\rightarrow{\bf C}$ defined by $\Xi_C^{\mu}(\Upsilon)=\Xi^{\mu}(\Upsilon)$ is countably additive and that $\Xi_C^{\mu}(\emptyset)=0$.  

Therefore $\Xi^{\mu}$ is a well defined complex matrix valued measure on $C_m^{o}$ for all $\mu=0,1,2,3$. It is not divergent either in the UV or the IR when its argument is a compact subset of $C_m^{o}$.

If, as we will show, the measure $\Xi^{\mu}$ is associated with a density $q\mapsto\Xi^{\mu}(q)$ which can be extended to a continuous function on $C_m$ then the object $\Phi^{\mu}$ given by Eq.~ \eqref{eq:Phi_def_t_channel} defines a measure given by
\begin{equation} \label{eq:Phi_Xi_relation}
\Phi^{\mu}(\Upsilon)=\int\Phi^{\mu}(q)\,dq=\frac{e^3}{(2\pi)^4}\int_{\Upsilon}\Xi^{\mu}(\frac{1}{2}q)\,dq=\frac{e^3}{\pi^4}\int_{\frac{1}{2}\Upsilon}\Xi^{\mu}(q)\,dq=\frac{e^3}{\pi^4}\Xi(\frac{1}{2}\Upsilon).
\end{equation}
In this case the measure $\Phi^{\mu}$ is a well defined complex matrix valued measure on $C_{2m}$. Thus, since $\{p^{\prime}+p:p^{\prime},p\in H_m\}=C_{2m}$, the measure $\Phi^{\mu}$ may be thought of as defining a mapping 
\begin{equation} \label{eq:Gamma_domain}
\Gamma^{\mu}:H_m\times H_m\rightarrow{\bf C}^{4\times4},\Gamma^{\mu}(p^{\prime},p)=\Phi(q),q=p^{\prime}+p\in C_{2m}.
\end{equation}

Using the gamma matrix contraction identities one can show that
\[ \gamma^{\nu}({\slas q}+m)\gamma^{\mu}({\slas q}+2{\slas k}-m)\gamma_{\nu}=-2{\slas q}\gamma^{\mu}{\slas q}-4{\slas k}\gamma^{\mu}{\slas q}+8mk^{\mu}+2m^2\gamma^{\mu}. \]
Therefore
\begin{equation}
\Xi^{\mu}(\Upsilon)=-\frac{\pi^2}{2}\int\chi_{\Upsilon}(q+k)({\slas q}\gamma^{\mu}{\slas q}+2{\slas k}\gamma^{\mu}{\slas q}-4mk^{\mu}-m^2\gamma^{\mu})(q.k)^{-1}\,\Omega_m(dq)\,\Omega_0^{+}(dk).
\end{equation}
We will show that $\Xi^{\mu}$ is a $K$ covariant ${\bf C}^{4\times4}$ valued measure on $C_m^{o}$ with one Lorentz index as follows.

Consider $\Xi_{\mu}=\eta_{\mu\nu}\Xi^{\nu}$. Then we have
\begin{align*}
\Xi_{\mu}(\kappa\Upsilon)=&-\frac{\pi^2}{2}\int\chi_{\kappa\Upsilon}(q+k)({\slas q}\gamma_{\mu}{\slas q}+2{\slas k}\gamma_{\mu}{\slas q}-4mk_{\mu}-m^2\gamma_{\mu})(q.k)^{-1}\,\Omega_m(dq)\,\Omega_0^{+}(dk)\\
=&-\frac{\pi^2}{2}\int\chi_{\Upsilon}(\kappa^{-1}q+\kappa^{-1}k)({\slas q}\gamma_{\mu}{\slas q}+2{\slas k}\gamma_{\mu}{\slas q}-4mk_{\mu}-m^2\gamma_{\mu})(q.k)^{-1}\,\Omega_m(dq)\,\Omega_0^{+}(dk)\\
=&-\frac{\pi^2}{2}\int\chi_{\Upsilon}(q+k)(\kappa{\slas q}\kappa^{-1}\gamma_{\mu}{\kappa\slas q}\kappa^{-1}+2\kappa{\slas k}\kappa^{-1}\gamma_{\mu}\kappa{\slas q}\kappa^{-1}-4m(\kappa k)_{\mu}-m^2\kappa\kappa^{-1}\gamma_{\mu}\kappa\kappa^{-1})\\
&((\kappa q).(\kappa k))^{-1}\,\Omega_m(dq)\,\Omega_0^{+}(dk)\\
=&-\frac{\pi^2}{2}\kappa\int\chi_{\Upsilon}(q+k)({\slas q}\Lambda^{-1\rho}{}_{\mu}\gamma_{\rho}{\slas q}+2{\slas k}\Lambda^{-1\rho}{}_{\mu}\gamma_{\rho}{\slas q}-4m\Lambda^{-1\rho}{}_{\mu}k_{\rho}-m^2\Lambda^{-1\rho}{}_{\mu}\gamma_{\rho})\\
&(q.k)^{-1}\,\Omega_m(dq)\,\Omega_0^{+}(dk)\,\kappa^{-1}\\
=&-\frac{\pi^2}{2}\Lambda^{-1\rho}{}_{\mu}\kappa\int\chi_{\Upsilon}(q+k)({\slas q}\gamma_{\rho}{\slas q}+2{\slas k}\gamma_{\rho}{\slas q}-4mk_{\rho}-m^2\gamma_{\rho})(q.k)^{-1}\,\Omega_m(dq)\,\Omega_0^{+}(dk)\,\kappa^{-1}\\
=&\Lambda^{-1\rho}{}_{\mu}\kappa\Xi_{\rho}(\Upsilon)\kappa^{-1},
\end{align*}
for all $\kappa\in K,\Upsilon\in{\mathcal B}_0(C_m^{o})$, where we have used the Lorentz invariance of $\Omega_m$ and $\Omega_0^{+}$, the fundamental intertwining property of the Feynman slash and the fact that
\begin{align*}
&(\kappa k)_{\mu}=(\Lambda k)_{\mu}=\eta_{\mu\nu}(\Lambda k)^{\nu}=\eta_{\mu\nu}\Lambda^{\nu}{}_{\sigma}k^{\sigma}=\eta_{\mu\nu}\Lambda^{\nu}{}_{\sigma}\eta^{\sigma\rho}k_{\rho}=\\
&(\eta\Lambda\eta)_{\mu}{}^{\rho}k_{\rho}=(\Lambda^{-1T})_{\mu}{}^{\rho}k_{\rho}=\Lambda^{-1\rho}{}_{\mu}k_{\rho}.
\end{align*}
Therefore, using Eq.~\eqref{eq:Lorentz_identity}
\begin{equation}
\Xi^{\mu}(\kappa\Upsilon)=\eta^{\mu\nu}\Xi_{\nu}(\kappa\Upsilon)=\eta^{\mu\nu}\Lambda^{-1\rho}{}_{\nu}\kappa\Xi_{\rho}(\Upsilon)\kappa^{-1}=\eta^{\rho\nu}\Lambda^{\mu}{}_{\nu}\kappa\Xi_{\rho}(\Upsilon)\kappa^{-1}=\Lambda^{\mu}{}_{\nu}\kappa\Xi^{\nu}(\Upsilon)\kappa^{-1},
\end{equation}
as required.

\subsection{Determination of the density $\Phi^{\mu}$}

We will now use the spectral calculus to compute the spectrum of $\Phi^{\mu}$ and then use this spectrum to compute the density for $\Phi^{\mu}$.
\begin{align*}
g^{\mu}(a,b,\epsilon)&=\Phi^{\mu}(\Upsilon(a,b,\epsilon))\\
&=\frac{e^3}{\pi^4}\Xi^{\mu}(\frac{1}{2}\Upsilon(a,b,\epsilon))\\
&=-\frac{e^3}{2\pi^2}\int\chi_{\frac{1}{2}\Upsilon(a,b,\epsilon)}(q+k)({\slas q}\gamma^{\mu}{\slas q}+2{\slas k}\gamma^{\mu}{\slas q}-4mk^{\mu}-m^2\gamma^{\mu})(q.k)^{-1}\,\Omega_m(dq)\,\Omega_0^{+}(dk)\\
&\approx-\frac{e^3}{2\pi^2}\int\chi_{(\frac{1}{2}a,\frac{1}{2}b)}(\omega_m({\vct q})+|{\vct k}|)\chi_{\frac{1}{2}B_{\epsilon}({\vct0})}({\vct q}+{\vct k})({\slas q}\gamma^{\mu}{\slas q}+2{\slas k}\gamma^{\mu}{\slas q}-4mk^{\mu}-m^2\gamma^{\mu})\\
&(q.k)^{-1}\frac{d{\vct q}}{\omega_m({\vct q})}\frac{d{\vct k}}{|{\vct k}|},\\
&\mbox{ where }q=(\omega_m({\vct q}),{\vct q}),k=(|{\vct k}|,{\vct k})\\
&=-\frac{e^3}{2\pi^2}\int\chi_{(\frac{1}{2}a,\frac{1}{2}b)}(\omega_m({\vct q})+|{\vct k}|)\chi_{\frac{1}{2}B_{\epsilon}({\vct0})-{\vct k}}({\vct q})({\slas q}\gamma^{\mu}{\slas q}+2{\slas k}\gamma^{\mu}{\slas q}-4mk^{\mu}-m^2\gamma^{\mu})\\
&(q.k)^{-1}\frac{d{\vct q}}{\omega_m({\vct q})}\frac{d{\vct k}}{|{\vct k}|},\\
&\mbox{ where }q=(\omega_m({\vct q}),{\vct q}),k=(|{\vct k}|,{\vct k})\\
&\approx-\frac{e^3}{2\pi^2}\int\chi_{(\frac{1}{2}a,\frac{1}{2}b)}(\omega_m({\vct k})+|{\vct k}|)({\slas q}\gamma^{\mu}{\slas q}+2{\slas k}\gamma^{\mu}{\slas q}-4mk^{\mu}-m^2\gamma^{\mu})(q.k)^{-1}\\
&\omega_m({\vct k})^{-1}|{\vct k}|^{-1}\,d{\vct k}\,(\frac{1}{6}\pi\epsilon^3),\\
&\mbox{ where }q=(\omega_m({\vct k}),{-\vct k}),k=(|{\vct k}|,{\vct k}).
\end{align*}
Therefore
\begin{align*}
g^{\mu}_a(b)&=\lim_{\epsilon\rightarrow0}\epsilon^{-3}g(a,b,\epsilon)\\
&=-\frac{e^3}{2\pi^2}\int\chi_{(\frac{1}{2}a,\frac{1}{2}b)}(\omega_m({\vct k})+|{\vct k}|)({\slas q}\gamma^{\mu}{\slas q}+2{\slas k}\gamma^{\mu}{\slas q}-4mk^{\mu}-m^2\gamma^{\mu})(q.k)^{-1}\\
&\omega_m({\vct k})^{-1}|{\vct k}|^{-1}\,d{\vct k}\,(\frac{1}{6}\pi),\\
&\mbox{ where }q=(\omega_m({\vct k}),-{\vct k}),k=(|{\vct k}|,{\vct k}).
\end{align*}
Let $b>a>m$. Then
\begin{align*}
\chi_{(a,b)}(\omega_m({\vct k})+|{\vct k}|)=1&\Leftrightarrow a<\omega_m({\vct k})+|{\vct k}|<b\\
&\Leftrightarrow a^2<m^2+{\vct k}^2+{\vct k}^2+2(m^2+{\vct k}^2)^{\frac{1}{2}}|{\vct k}|<b^2\\
&\Leftrightarrow(a^2-m^2-2X)^2<4(m^2+X)X<(b^2-m^2-2X)^2\\
&\Leftrightarrow(a^2-m^2)^2+4X^2-4(a^2-m^2)X<4m^2X+4X^2<\\
&(b^2-m^2)^2+4X^2-4(b^2-m^2)X\\
&\Leftrightarrow(a^2-m^2)^2-4a^2X<0<(b^2-m^2)^2-4b^2X\\
&\Leftrightarrow Z(a)^2<X<Z(b)^2,
\end{align*}
where $X={\vct k}^2$ and
\begin{equation}
Z(s)=\frac{s^2-m^2}{2s}.
\end{equation}
Therefore, using spherical polar coordinates, we have that
\begin{align*}
g^{\mu}_a(b)&=-\frac{e^3}{2\pi^2}\int_{s=Z(\frac{1}{2}a)}^{Z(\frac{1}{2}b)}\int_{\theta=0}^{\pi}\int_{\phi=0}^{2\pi}({\slas q}\gamma^{\mu}{\slas q}+2{\slas k}\gamma^{\mu}{\slas q}-4mk^{\mu}-m^2\gamma^{\mu})(\omega_m(s)s+s^2)^{-1}\\
&\omega_m(s)^{-1}s^{-1}s^2\sin(\theta)\,d\phi\,d\theta\,ds\,(\frac{1}{6}\pi),\\
&\mbox{ where }q=(\omega_m(s),-{\vct k}),k=(s,{\vct k}),\omega_m(s)=(m^2+s^2)^{\frac{1}{2}},\\
&{\vct k}=s(\sin(\theta)\cos(\phi),\sin(\theta)\sin(\phi),\cos(\theta)).
\end{align*}
Consider
\begin{equation}
X=\int_{\theta=0}^{\pi}\int_{\phi=0}^{2\pi}{\slas q}\gamma^{\mu}{\slas q}\sin(\theta)\,d\phi\,d\theta.
\end{equation}
Then
\begin{align*}
X&=\int_{\theta=0}^{\pi}\int_{\phi=0}^{2\pi}(\omega_m(s)\gamma^0-s\sin(\theta)\cos(\phi)\gamma^1-s\sin(\theta)\sin(\phi)\gamma^2-s\cos(\theta)\gamma^3)\gamma^{\mu}\\
&(\omega_m(s)\gamma^0-s\sin(\theta)\cos(\phi)\gamma^1-s\sin(\theta)\sin(\phi)\gamma^2-s\cos(\theta)\gamma^3)\sin(\theta)\,d\phi\,d\theta\\
&=4\pi(m^2+s^2)\gamma^0\gamma^{\mu}\gamma^0+\int_{\theta=0}^{\pi}\int_{\phi=0}^{2\pi}(s^2\sin^2(\theta)\cos^2(\phi)\gamma^1\gamma^{\mu}\gamma^1+s^2\sin^2(\theta)\sin^2(\phi)\gamma^2\gamma^{\mu}\gamma^2+\\
&s^2\cos^2(\theta)\gamma^3\gamma^{\mu}\gamma^3)\sin(\theta)\,d\phi\,d\theta\\
&=4\pi(m^2+s^2)\gamma^0\gamma^{\mu}\gamma^0+Y,
\end{align*}
say (all other terms vanish).\\
When $\mu=0$
\begin{align*}
Y&=\int_{\theta=0}^{\pi}\int_{\phi=0}^{2\pi}(s^2\sin^2(\theta)\cos^2(\phi)\gamma^0+s^2\sin^2(\theta)\sin^2(\phi)\gamma^0+\\
&s^2\cos^2(\theta)\gamma^0)\sin(\theta)\,d\phi\,d\theta\\
&=4\pi s^2\gamma^0.
\end{align*}
When $\mu=1$
\begin{align*}
Y&=\int_{\theta=0}^{\pi}\int_{\phi=0}^{2\pi}(-s^2\sin^2(\theta)\cos^2(\phi)\gamma^1+s^2\sin^2(\theta)\sin^2(\phi)\gamma^1+\\
&s^2\cos^2(\theta)\gamma^1)\sin(\theta)\,d\phi\,d\theta\\
&=2\pi s^2\gamma^1\int_{u=-1}^1u^2\,du\\
&=\frac{4}{3}\pi s^2\gamma^1.
\end{align*}
When $\mu=2$
\begin{align*}
Y&=\int_{\theta=0}^{\pi}\int_{\phi=0}^{2\pi}(s^2\sin^2(\theta)\cos^2(\phi)\gamma^2-s^2\sin^2(\theta)\sin^2(\phi)\gamma^2+\\
&s^2\cos^2(\theta)\gamma^2)\sin(\theta)\,d\phi\,d\theta\\
&=2\pi s^2\gamma^2\int_{u=-1}^1u^2\,du\\
&=\frac{4}{3}\pi s^2\gamma^2.
\end{align*}
When $\mu=3$
\begin{align*}
Y&=\int_{\theta=0}^{\pi}\int_{\phi=0}^{2\pi}(s^2\sin^2(\theta)\cos^2(\phi)\gamma^{3}+s^2\sin^2(\theta)\sin^2(\phi)\gamma^{3}-\\
&s^2\cos^2(\theta)\gamma^3)\sin(\theta)\,d\phi\,d\theta\\
&=2\pi s^2\int_{\theta=0}^{\pi}(\sin^2(\theta)-\cos^2(\theta))\sin(\theta)\,d\theta\,\gamma^{3}\\
&=\frac{4}{3}\pi s^2\gamma^3.
\end{align*}
Therefore
\begin{equation}\label{eq:phi_theta_integral}
X=4\pi(m^2+s^2)\gamma^0\gamma^{\mu}\gamma^0+\frac{4}{3}\pi s^2\gamma^{\mu}+\frac{8}{3}\pi s^2\gamma^{0}\eta^{\mu0}.
\end{equation}
Now consider
\begin{equation}
X=\int_{\theta=0}^{\pi}\int_{\phi=0}^{2\pi}{\slas k}\gamma^{\mu}{\slas q}\sin(\theta)\,d\phi\,d\theta.
\end{equation}
Then
\begin{align*}
X&=\int_{\theta=0}^{\pi}\int_{\phi=0}^{2\pi}(s\gamma^0+s\sin(\theta)\cos(\phi)\gamma^1+s\sin(\theta)\sin(\phi)\gamma^2+s\cos(\theta)\gamma^3)\gamma^{\mu}\\
&(\omega_m(s)\gamma^0-s\sin(\theta)\cos(\phi)\gamma^1-s\sin(\theta)\sin(\phi)\gamma^2-s\cos(\theta)\gamma^3)\sin(\theta)\,d\phi\,d\theta\\
&=4\pi s\omega_m(s)\gamma^0\gamma^{\mu}\gamma^0-\int_{\theta=0}^{\pi}\int_{\phi=0}^{2\pi}(s^2\sin^2(\theta)\cos^2(\phi)\gamma^1\gamma^{\mu}\gamma^1+s^2\sin^2(\theta)\sin^2(\phi)\gamma^2\gamma^{\mu}\gamma^2+\\
&s^2\cos^2(\theta)\gamma^3\gamma^{\mu}\gamma^3)\sin(\theta)\,d\phi\,d\theta,
\end{align*}
and it is straightforward, arguing as above, to show that this evaluates to
\begin{equation}
X=4\pi s\omega_m(s)\gamma^0\gamma^{\mu}\gamma^0-\frac{4}{3}\pi s^2\gamma^{\mu}-\frac{8}{3}\pi s^2\gamma^{0}\eta^{\mu0}.
\end{equation}
Now consider
\begin{equation}
X=\int_{\theta=0}^{\pi}\int_{\phi=0}^{2\pi}k^{\mu}\sin(\theta)\,d\phi\,d\theta.
\end{equation}
Then
\begin{align*}
X&=\int_{\theta=0}^{\pi}\int_{\phi=0}^{2\pi}(s,s\sin(\theta)\cos(\phi),s\sin(\theta)\sin(\phi),s\cos(\theta))^{\mu}\sin(\theta)\,d\phi\,d\theta\\
&=4\pi s\eta^{\mu0}.
\end{align*}
Now putting together all the computations for the various quantities ``X" into the equation that we derived above for $g^{\mu}_a$ we obtain
\begin{align*}
g^{\mu}_a(b)=&-\frac{e^3}{2\pi^2}\int_{s=Z(\frac{1}{2}a)}^{Z(\frac{1}{2}b)}[4\pi(m^2+s^2)\gamma^0\gamma^{\mu}\gamma^0+\frac{4}{3}\pi s^2\gamma^{\mu}+\frac{8}{3}\pi s^2\gamma^{0}\eta^{\mu0}+2(4\pi s\omega_m(s)\gamma^0\gamma^{\mu}\gamma^0-\\
&\frac{4}{3}\pi s^2\gamma^{\mu}-\frac{8}{3}\pi s^2\gamma^{0}\eta^{\mu0})-4m(4\pi s\eta^{\mu0})-4\pi m^2\gamma^{\mu}]\\
&(\omega_m(s)s+s^2)^{-1}\omega_m(s)^{-1}s^{-1}s^2\,ds\,(\frac{1}{6}\pi).
\end{align*}
Therefore, using the Leibniz integral rule, the spectrum $M^{\mu}$ of $\Phi^{\mu}$ is
\begin{align*}
M^{\mu}(b)=&\frac{3}{4\pi}g_a^{\mu\prime}(b)\\
=&-\frac{e^3}{16\pi^2}[(4\pi(m^2+s^2)+8\pi s\omega_m(s))\gamma^0\gamma^{\mu}\gamma^0-(4\pi m^2+\frac{4}{3}\pi s^2)\gamma^{\mu}-\frac{8}{3}\pi s^2\gamma^{0}\eta^{\mu0}-\\
&\left.16\pi ms\eta^{\mu0})(\omega_m(s)+s)^{-1}\omega_m(s)^{-1}]\frac{ }{ }\right|_{s=Z(\frac{1}{2}b)}\frac{1}{2}Z^{\prime}(\frac{1}{2}b).
\end{align*}
Thus
\begin{equation}\label{eq:t_vertex_spectrum}
M^{\mu}(b)=f_1(b)\gamma^{\mu}+f_2(b)\gamma^0\gamma^{\mu}\gamma^0+f_3(b)\eta^{\mu0}+f_4(b)\gamma^{0}\eta^{\mu0},
\end{equation}
where
\begin{align*}
f_1(b)&=\frac{e^3}{8\pi}[(m^2+\frac{1}{3}s^2)(\omega_m(s)+s)^{-1}\\
&\omega_m(s)^{-1}]\left.\frac{\mbox{}}{\mbox{}}\right|_{s=Z(\frac{1}{2}b)}Z^{\prime}(\frac{1}{2}b),\\
f_2(b)&=-\frac{e^3}{8\pi}[(m^2+s^2+2s\omega_m(s))(\omega_m(s)+s)^{-1}\\
&\omega_m(s)^{-1}]\left.\frac{\mbox{}}{\mbox{}}\right|_{s=Z(\frac{1}{2}b)}Z^{\prime}(\frac{1}{2}b),\\
f_3(b)&=\frac{e^3}{2\pi}[ms(\omega_m(s)+s)^{-1}\\
&\omega_m(s)^{-1}]\left.\frac{\mbox{}}{\mbox{}}\right|_{s=Z(\frac{1}{2}b)}Z^{\prime}(\frac{1}{2}b),\\
f_4(b)&=\frac{e^3}{12\pi}[s^2(\omega_m(s)+s)^{-1}\\
&\omega_m(s)^{-1}]\left.\frac{\mbox{}}{\mbox{}}\right|_{s=Z(\frac{1}{2}b)}Z^{\prime}(\frac{1}{2}b).
\end{align*}
Hence the spectrum $M^{\mu}(b)$ of $\Phi^{\mu}$ is a continuous (in fact analytic) function of $b$ for all $b>2m$.

Now, since
\[ \gamma^0\gamma^{\mu}\gamma^0=\gamma^0(2\eta^{\mu0}-\gamma^0\gamma^{\mu})=2\eta^{\mu0}\gamma^0-\gamma^{\mu}, \]
the spectrum of $\Phi^{\mu}$ can be written as
\begin{equation}\label{eq:M_mu_spectrum}
M^{\mu}(b)=(f_1(b)-f_2(b))\gamma^{\mu}+f_3(b)\eta^{\mu0}+(f_4(b)+2f_2(b))\gamma^0\eta^{\mu0},
\end{equation} 
and has the form of Eq.~\eqref{eq:spectrum} for the spectrum for the canonical $K$ covariant measure on Minkowski space with one Lorentz index defined on $C_{2m}^{o}$. This spectrum extends continuously to a function on $[2m,\infty)$ defined by the equation~\ref{eq:t_vertex_spectrum}. 

Comparing Eq.~\eqref{eq:M_mu_spectrum} for the spectrum of $\Phi^{\mu}$ with the canonical spectral form given by Eq.~\eqref{eq:spectrum} we can read off the spectral functions $\sigma_1,\ldots,\sigma_4$ as follows.
\begin{align*}
&b^{-1}\sigma_1(b)=f_1(b)-f_2(b),\\
&\sigma_3(b)=f_3(b),\\
&b\sigma_4(b)=f_4(b)+2f_2(b),
\end{align*}
for all $b\geq2m$, while $\sigma_2$ vanishes. Using Eq.~\eqref{eq:density} the density corresponding to $\Phi^{\mu}$ is
\begin{align}
\Phi^{\mu}(q)=Q^{-1}(\sigma_1(Q)\gamma^{\mu}+\sigma_3(Q)q^{\mu}+\sigma_4(Q){\slas q}q^{\mu}),
\end{align}
where $Q=(q^2)^{\frac{1}{2}}$. 
Thus
\begin{align}
\Phi^{\mu}(q)&=(f_1(Q)-f_2(Q))\gamma^{\mu}+Q^{-1}f_3(Q)q^{\mu}+Q^{-2}(f_4(Q)+2f_2(Q)){\slas q}q^{\mu}.\label{eq:Phi_mu}
\end{align}
where $Q=(q^2)^{\frac{1}{2}},q^2\geq4m^2$.

\section{The anomalous magnetic moment of the electron\label{section:anom_mag_mom}}

The calculation of the anomalous magnetic moment of the electron, first carried out at one-loop level by Schwinger \cite{Schwinger,Schwinger1}, is a result of great importance in QFT \cite{Kronig,Darwin,Kusch,Levine,Carroll,Commins}.

To determine the LO contribution to the anomalous magnetic moment of the electron we determine the effect of the vertex correction relative to the tree level diagram. The Feynman subamplitude ${\mathcal M}_{\mbox{tr}}^{\mu}$ associated with the electron component of the tree level diagram for the interaction of an electron with an external electromagnetic field is given by
\begin{equation}
{\mathcal M}^{\mu}_{\mbox{tr}}(p^{\prime},p)=\overline{u}(p^{\prime})\Phi_{\mbox{tr}}^{\mu}(q)u(p),
\end{equation}
where
\begin{equation}
\Phi_{\mbox{tr}}^{\mu}(q)=ie\gamma^{\mu}.
\end{equation}
The vertex correction makes a contribution of
\begin{equation}
{\mathcal M}_{\mbox{v}}^{\mu}(p^{\prime},p)=\overline{u}(p^{\prime})\Gamma^{\mu}(p^{\prime},p)u(p)=\overline{u}(p^{\prime})\Phi^{\mu}(q)u(p).
\end{equation} 
When sandwiched between Dirac spinors the term in $\Phi^{\mu}$ involving ${\slas q}q^{\mu}$ vanishes. Therefore it can be dropped.
 
Now, since we have carried out a flip operation, $q=p^{\prime}-Tp=p^{\prime}+Pp$ where $P:{\bf R}^4\rightarrow{\bf R}^4$ is the parity operator defined by
\begin{equation}
Pp=P(p^0,{\vct p})=(p^0,-{\vct p}).
\end{equation}
It follows that $q\in C_{2m}$.

The measure on the mass shell $q\in H_{m^{\prime}}$ for $m^{\prime}\geq2m$ associated with the vertex correction is
\begin{equation}
\Upsilon\mapsto\int_{\Upsilon}\Phi^{\mu}(q)\,\Omega_{m^{\prime}}(dq).
\end{equation}
In particular at low energy $m^{\prime}=2m$ when $q\in H_{2m}$, since
\begin{align*}
&Z(m)=0,\\
&Z^{\prime}(m)=1,\\
&f_1(2m)=\frac{e^3}{8\pi},\\
&f_2(2m)=-\frac{e^3}{8\pi},\\
&f_3(2m)=0,\\
&f_4(2m)=0,
\end{align*}
we have
\[ \Phi^{\mu}(q)=(f_1(2m)-f_2(2m))\gamma^{\mu}=\frac{e^3}{4\pi}\gamma^{\mu}. \]
Therefore
\begin{equation}
\int_{\Upsilon}\Phi^{\mu}(q)\,\Omega_{2m}(dq)=\frac{e^3}{4\pi}\gamma^{\mu}\int\chi_{\Upsilon}((\omega_{2m}({\vct q}),{\vct q}))\frac{d{\vct q}}{\omega_{2m}({\vct q})}.
\end{equation}
If $\Upsilon_{\mbox{\scriptsize small}}=(0,b)\times S$ where $b>2m$ and $S\subset{\bf R}^3$ ($S\in{\mathcal B}_0({\bf R}^3)$) is such that $S$ corresponds to low momenta, i.e. $|{\vct q}|$ is small for all ${\vct q}\in S$ then
\[ \int_{\Upsilon_{\mbox{\scriptsize small}}}\Phi^{\mu}(q)\,\Omega_{2m}(dq)\approx\frac{e^3}{4\pi}\gamma^{\mu}\frac{1}{2m}{\mathcal L}(S), \]
where 
\[ {\mathcal L}(S)=\int_Sd{\vct q}. \]
The automorphism group of the Feynman diagram of Figure~\ref{figure:QFT_vertex} has order 2. Therefore the diagram has a a symmetry factor of $\frac{1}{2}$ associated with it. Therefore, in fact, we have
\[ \int_{\Upsilon_{\mbox{\scriptsize small}}}\Phi^{\mu}(q)\,\Omega_{2m}(dq)\approx\frac{e^3}{8\pi}\gamma^{\mu}\frac{1}{2m}{\mathcal L}(S). \]
Putting $\Phi_{\mbox{tr}}$ on the mass shell $H_{2m}$ and using the result given by Eq.~\eqref{eq:ansatz} we obtain
\begin{align*}
\int_{\Upsilon}\frac{1}{q^2-(2m)^2+i\epsilon}\Phi^{\mu}_{\mbox{tr}}(q)\,dq=&-i\pi\int_{\Upsilon}\Phi_{\mbox{tr}}^{\mu}(q)\,\Omega_{2m}(dq)=-i\pi(ie\gamma^{\mu})\int_{\Upsilon}\Omega_{2m}(dq)\\
&=\pi e\gamma^{\mu}\int\chi_{\Upsilon}((\omega_{2m}({\vct q}),{\vct q}))\frac{d{\vct q}}{\omega_{2m}({\vct q})}.
\end{align*}
For low momenta this evaluates to
\[ \int_{\Upsilon_{\mbox{\scriptsize small}}}\frac{1}{q^2-(2m)^2+i\epsilon}\Phi^{\mu}_{\mbox{tr}}(q)\,dq\approx\pi e\gamma^{\mu}\frac{1}{2m}{\mathcal L}(S). \]
Adding these two contributions the total measure associated with low energy and low momenta for the tree + vertex correction process is
\begin{align*}
\mbox{total}(S)&=\pi e\gamma^{\mu}\frac{1}{2m}{\mathcal L}(S)+\frac{e^3}{8\pi}\gamma^{\mu}\frac{1}{2m}{\mathcal L}(S)\\
&=\pi e\gamma^{\mu}\frac{1}{2m}{\mathcal L}(S)(1+\frac{e^2}{8\pi^2})\\
&=(1+\frac{e^2}{8\pi^2})\mbox{tree}(S)\\
&=(1+\frac{\alpha}{2\pi})\mbox{tree}(S),
\end{align*}
where $\alpha=\frac{e^2}{4\pi}$ is the fine structure constant,
and we have therefore derived the well known result of Schwinger.

\section{The vertex function in the s channel\label{section:vertex_function_s}}

\begin{figure} 
\centering
\includegraphics[width=6cm]{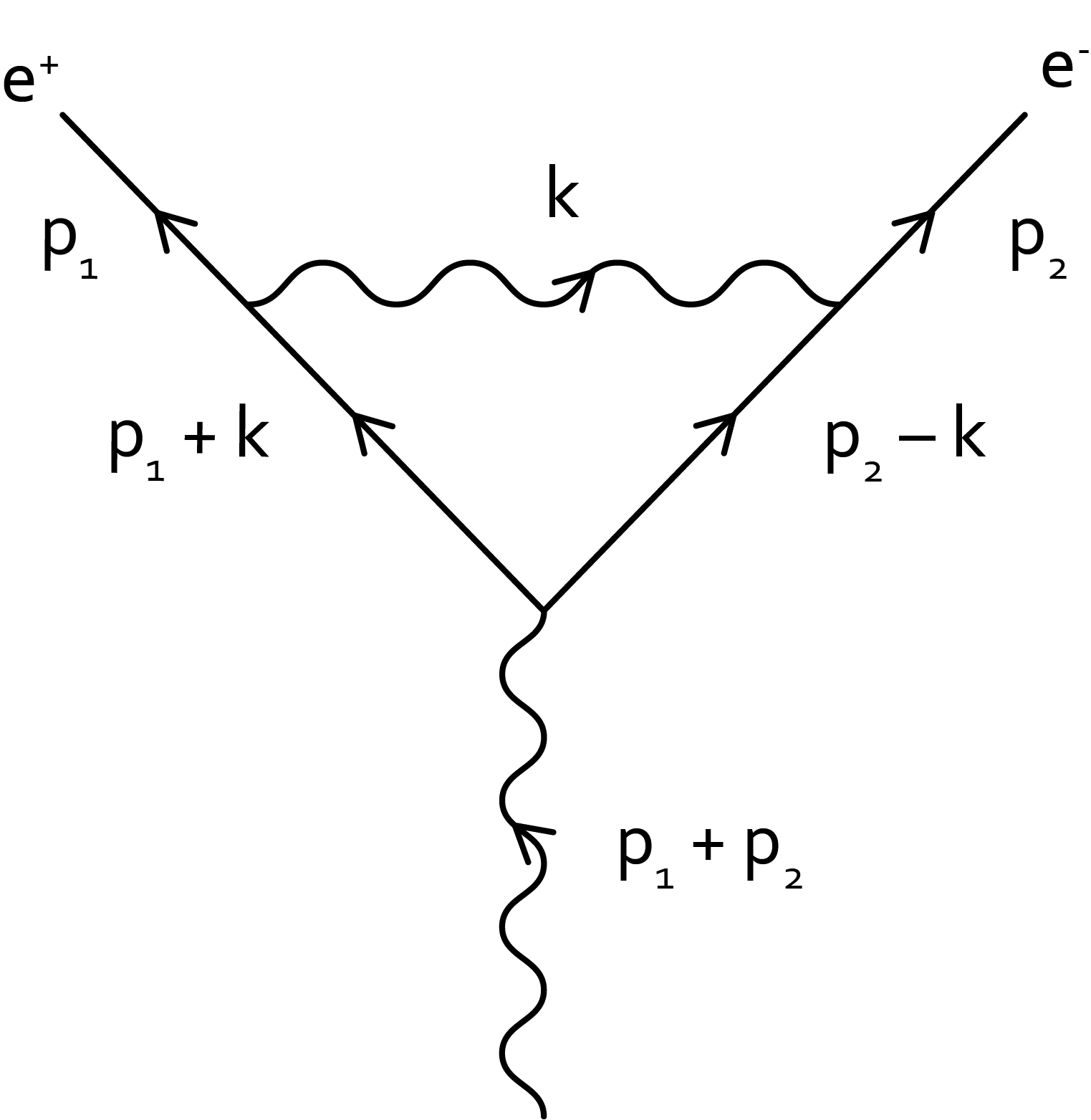}
\caption{Feynman diagram for the vertex function in the $s$ channel} \label{figure:s_channel_1}
\end{figure}

Consider the Feynman diagram given by Figure \ref{figure:s_channel_1} describing the QFT vertex function in the s channel. The diagram describes the QFT vertex function associated with a positively charged particle of momentum $p_1$, charge $e$ and mass $m$ and a negatively charged particle of momentum $p_2$, charge $-e$ and mass $m$ in the s channel, e.g. electron-positron pair production or muon pair production. We view antiparticles as being particles of negative energy (this is equivalent to viewing them as being particles travelling backwards in time).  
Then $p_1$ is on the mass shell $H_m^{-}$ and $p_2$ is on the mass shell $H_m^{+}$.  Thus $Tp_1,p_2\in H_m$.
Therefore, using the CM frame, there is, by a well known theorem, a ${\vct p}\in{\bf R}^3$ such that
\begin{equation}
Tp_1=(E,-{\vct p}),p_2=(E,{\vct p}),
\end{equation}
where $E=\omega_m({\vct p})$.
Therefore
\begin{equation}
p_1=(-E,-{\vct p}),p_2=(E,{\vct p}).
\end{equation}
Let 
\begin{equation}
q=p_2-p_1. \label{eq:q_def}
\end{equation}
Then
\begin{equation}
p_1=-\frac{1}{2}q,p_2=\frac{1}{2}q,q=2(E,{\vct p}). \label{eq:CM_p_q}
\end{equation}

Applying the Feynman rules the vertex function in the s channel is described by the Feynman integral
\begin{equation} \label{eq:vertex_modification_integral_2}
\Gamma^{\mu}=\int\frac{d^4k}{(2\pi)^4}iD_{\rho\sigma}(k)ie\gamma^{\rho}iS(p_2-k)i\gamma^{\mu}iS(p_1+k)ie\gamma^{\sigma},
\end{equation}
where, for convenience, we do not include here the factor of $e$ associated with the central vertex.

\subsection{Determination of a formal representation for the s channel vertex function as a ${\bf C}^{4\times4}$ valued measure with one index}

From Eq.~\eqref{eq:vertex_modification_integral_2} 
\begin{equation}\label{eq:original_vertex_equation_s_channel}
\Gamma^{\mu}(p_1,p_2)=\frac{e^2}{(2\pi)^4}\int\frac{1}{k^2+i\epsilon}\gamma^{\nu}\frac{1}{{\slas p_2}-{\slas k}-m+i\epsilon}\gamma^{\mu}\frac{1}{{\slas p}_1+{\slas k}-m+i\epsilon}\gamma_{\nu}\,dk.
\end{equation}

Therefore, from Eqns.~\eqref{eq:original_vertex_equation_s_channel} and \eqref{eq:CM_p_q},
\begin{align*}
\Gamma^{\mu}(p_1,p_2)=&\frac{e^2}{(2\pi)^4}\int\frac{1}{k^2+i\epsilon}\gamma^{\nu}\frac{1}{\frac{{\slas q}}{2}-{\slas k}-m+i\epsilon}\gamma^{\mu}\frac{1}{-\frac{{\slas q}}{2}+{\slas k}-m+i\epsilon}\gamma_{\nu}\,dk\\
=&\frac{e^2}{(2\pi)^4}\int\frac{1}{k^2+i\epsilon}\gamma^{\nu}\frac{2}{{\slas q}-2{\slas k}-2m+i\epsilon}\gamma^{\mu}\frac{2}{-{\slas q}+2{\slas k}-2m+i\epsilon}\gamma_{\nu}\,dk\\
=&\frac{e^2}{(2\pi)^4}\int\frac{1}{(\frac{k}{2})^2+i\epsilon}\gamma^{\nu}\frac{2}{{\slas q}-{\slas k}-2m+i\epsilon}\gamma^{\mu}\frac{2}{-{\slas q}+{\slas k}-2m+i\epsilon}\gamma_{\nu}(\frac{1}{16})\,dk\\
=&\frac{e^2}{(2\pi)^4}\int\frac{1}{k^2+i\epsilon}\gamma^{\nu}\frac{1}{{\slas q}-{\slas k}-2m+i\epsilon}\gamma^{\mu}\frac{1}{-{\slas q}+{\slas k}-2m+i\epsilon}\gamma_{\nu}\,dk\\
=&\Phi_{2m}^{\mu}(q),
\end{align*}
where, for $m>0$,
\begin{equation}
\Phi_m(q)=\frac{e^2}{(2\pi)^4}\Xi_m^{\mu}(q), \label{eq:Phi_def}
\end{equation}
in which, 
\begin{equation} \label{eq:Xi_m_def}
\Xi_m^{\mu}(q)=\int\frac{1}{k^2+i\epsilon}\gamma^{\nu}\frac{1}{{\slas q}-{\slas k}-m+i\epsilon}\gamma^{\mu}\frac{1}{-{\slas q}+{\slas k}-m+i\epsilon}\gamma_{\nu}\,dk.
\end{equation}

The integral defining $\Xi_m^{\mu}(q)$ does not converge for any $q\in{\bf R}^4$ so regularization is required. We will use covariant spectral regularization. 

We will show, by a formal argument, that $\Xi_m^{\mu}$ can be interpreted as a matrix valued measure with one Lorentz index as follows. 
Suppose (``pretend") that $\Xi_m^{\mu}$ existed as a ${\bf C}^{4\times4}$ valued function on Minkowski space. Then it is associated with a measure, which we shall also denote as $\Xi_m^{\mu}$, defined by
\begin{align*}
\Xi_m^{\mu}(\Upsilon)&=\int_{\Upsilon}\Xi_m^{\mu}(q)\,dq\\
&=\int\chi_{\Upsilon}(q)\Xi_m^{\mu}(q)\,dq\\
&=\int\chi_{\Upsilon}(q)(\int\frac{1}{k^2+i\epsilon}\gamma^{\nu}\frac{1}{{\slas q}-{\slas k}-m+i\epsilon}\gamma^{\mu}\frac{1}{-{\slas q}+{\slas k}-m+i\epsilon}\gamma_{\nu}\,dk)\,dq\\
&=\int\chi_{\Upsilon}(q)\frac{1}{k^2+i\epsilon}\gamma^{\nu}\frac{1}{{\slas q}-{\slas k}-m+i\epsilon}\gamma^{\mu}\frac{1}{-{\slas q}+{\slas k}-m+i\epsilon}\gamma_{\nu}\,dk\,dq\\
&"="\int\chi_{\Upsilon}(q)\frac{1}{k^2+i\epsilon}\gamma^{\nu}\frac{1}{{\slas q}-{\slas k}-m+i\epsilon}\gamma^{\mu}\frac{1}{-{\slas q}+{\slas k}-m+i\epsilon}\gamma_{\nu}\,dq\,dk\\
&=\int\chi_{\Upsilon}(q+k)\frac{1}{k^2+i\epsilon}\gamma^{\nu}\frac{1}{{\slas q}-m+i\epsilon}\gamma^{\mu}\frac{1}{-{\slas q}-m+i\epsilon}\gamma_{\nu}\,dq\,dk,
\end{align*}
where we have enclosed the = sign in quotes at the only unjustified formal step (interchange of the order of integration). 
Therefore we write
\begin{equation}
\Xi_m^{\mu}(\Upsilon)=-\int\chi_{\Upsilon}(q+k)\frac{1}{k^2+i\epsilon}\left(\frac{1}{q^2-m^2+i\epsilon}\right)^2\gamma^{\nu}({\slas q}+m)\gamma^{\mu}({\slas q}-m)\gamma_{\nu}\,dq\,dk. \label{eq:Xi_m_mu_1}
 \end{equation}
We will see shortly that the RHS of this equation can be given a well defined interpretation as a $K$ covariant matrix valued measure on Minkowski space with one Lorentz index. 

Note that there must be one or more unjustified formal steps in the argument that we have used because the ``function" defined by Eq.~\eqref{eq:Xi_m_def} is not well defined and one cannot go from something not well defined to something well defined by a rigorous argument consisting of a sequence of equalities.

\subsection{Determination of the distribution representing $(q^2-m^2+i\epsilon)^{-2}$}

Now we need to give a natural interpretation for the object
\begin{equation}\label{eq:f_def}
f(q)=\left(\frac{1}{q^2-m^2+i\epsilon}\right)^2.
\end{equation}

We have, omitting the $i\epsilon$,
\begin{equation}
f(q)=\frac{1}{((q^0)^2-\omega_m({\vct q})^2)^2}.
\end{equation}
Let
\begin{equation}
g(q)=\frac{1}{q^2-m^2}=\frac{1}{(q^0)^2-\omega_m({\vct q})^2}.
\end{equation}
Then
\[ (\partial_0g)(q)=-\frac{1}{((q^0)^2-\omega_m({\vct q}))^2}(2q^0)=-2q^0f(q). \]
Thus
\begin{equation}
f(q)=-\frac{1}{2q^0}(\partial_0g)(q). 
\end{equation}
For on shell values of its argument $g$ is best thought of as a distribution (measure on $H_m$). Similarly, for on shell values of its argument, $f$ is best thought of as a distribution.  We have, for any Schwartz function $\psi\in{\mathcal S}({\bf R}^4,{\bf C})$,
\begin{align*}
<f,\psi>=&\int f(q)\psi(q)\,dq\\
=&-\frac{1}{2}\int\psi(q)\frac{1}{q^0}(\partial_0 g)(q)\,dq\\
=&\frac{1}{2}\int\frac{\partial}{\partial q^0}(\psi(q)\frac{1}{q^0})g(q)dq\\
=&\frac{1}{2}\int((\partial_0\psi)(q)\frac{1}{q^0}-\psi(q)\frac{1}{(q^0)^2})g(q)dq\\
=&-\frac{1}{2}\pi i\int((\partial_0\psi)(q)\frac{1}{q^0}-\psi(q)\frac{1}{(q^0)^2})\,\Omega_m^{\pm}(dq),
\end{align*}
where we have used the result \cite{Symmetry,IJMPA} $g(q)dq\rightarrow-i\pi\Omega_m^{\pm}(dq)$.

For $m\geq0$ and $n\in\{1,2,\ldots\}$ define the measure $\Omega_m^{(n)\pm}$ by
\begin{equation}
\Omega_m^{(n)\pm}(\Upsilon)=\int_{\pi(H_{\pm m}\cap\Upsilon)}\omega_m({\vct q})^{-n}\,d{\vct q},
\end{equation}
where $\pi:{\bf R}^4\rightarrow{\bf R}^3$ is the natural projection defined by $\pi(q^0,{\vct q})={\vct q}$. One can equivalently define these measures by their effect on (measurable) functions as follows.
\begin{equation}
\int\psi(q)\,\Omega^{(n)\pm}(dq)=\int_{{\bf R}^3}\psi(\pm\omega_m({\vct q}),{\vct q})\omega_m({\vct q})^{-n}\,d{\vct q}.
\end{equation}
$\Omega_m^{(n)\pm}$ are measures concentrated on $H_m^{\pm}$ and, as is well known, $\Omega_m^{(1)\pm}=\Omega_m^{\pm}$ is Lorentz invariant.

We thus have
\begin{align*}
<f,\psi>=&-\frac{1}{2}\pi i(\int(\partial_0\psi)(q)\,\Omega_m^{(2)\pm}(dq)-\int\psi(q)\,\Omega_m^{(3)\pm}(dq))\\
=&\frac{1}{2}\pi i(\int \psi(q)\,(\partial_0\Omega_m^{(2)\pm})(dq)+\int\psi(q)\,\Omega_m^{(3)\pm}(dq)).
\end{align*}
Therefore we have the simple result that
\begin{equation}\label{eq:f_answer}
f=\frac{1}{2}\pi i(\Omega_m^{(3)\pm}+\partial_0\Omega_m^{(2)\pm}).
\end{equation}

Now by examination of the original formula given by Eq.~\eqref{eq:f_def} defining $f$ we may say that $f$ is manifestly Lorentz invariant. Therefore we must insist that, if we have calculated correctly, the object given by Eq.~\eqref{eq:f_answer} is Lorentz invariant. (This can be verified computationally).

\subsection{The causal $K$ covariant measure with one Lorentz index associated with $\Xi_m^{\mu}$}

We have, also, that \cite{Symmetry,IJMPA}
\begin{equation}\label{eq:prop_0_answer}
\frac{1}{k^2+ i\epsilon}\rightarrow-i\pi\Omega_0^{\pm}.
\end{equation}

Given Eqns.~\eqref{eq:Xi_m_mu_1}, \eqref{eq:f_def}, \eqref{eq:f_answer} and \eqref{eq:prop_0_answer}  we may consider the following object associated with $\Xi_m^{\mu}$
\begin{align*}
\Xi_m^{\mu}(\Upsilon)=&-\frac{1}{2}\pi^2(\int\chi_{\Upsilon}(q+k)\gamma^{\nu}({\slas q}+m)\gamma^{\mu}({\slas q}-m)\gamma_{\nu}\,(\partial_0\Omega_m^{(2)+})(dq)\,\Omega_0^{+}(dk)\\
+&\int\chi_{\Upsilon}(q+k)\gamma^{\nu}({\slas q}+m)\gamma^{\mu}({\slas q}-m)\gamma_{\nu}\,\Omega_m^{(3)+}(dq)\,\Omega_0^{+}(dk)).
\end{align*}
This means that $\Xi_m^{\mu}$ is the tempered distribution defined by
\begin{align}
<\Xi_m^{\mu},\psi>=&<\Xi_{1,m}^{\mu},\psi>+<\Xi_{2,m}^{\mu},\psi>, \label{eq:Xi_distribution}
\end{align}
where
\begin{equation}
<\Xi_{1,m}^{\mu},\psi>=\frac{1}{2}\pi^2\int\frac{\partial}{\partial q^0}(\psi(q+k)\gamma^{\nu}({\slas q}+m)\gamma^{\mu}({\slas q}-m)\gamma_{\nu})\,\Omega_m^{(2)+}(dq)\,\Omega_0^{+}(dk),
\end{equation}
and
\begin{equation}
<\Xi_{2,m}^{\mu},\psi>=-\frac{1}{2}\pi^2\int\psi(q+k)\gamma^{\nu}({\slas q}+m)\gamma^{\mu}({\slas q}-m)\gamma_{\nu}\,\Omega_m^{(3)+}(dq)\,\Omega_0^{+}(dk),
\end{equation}
for $\psi\in{\mathcal S}({\bf R}^4,{\bf C})$, the Schwartz space.

$\partial_0\Omega_m^{(2)+}$ is defined on the space ${\mathcal S}({\bf R}^4,{\bf C})$ by
\begin{equation}\label{eq:partial_Omega_def}
<\partial_0\Omega_m^{(2)+},\psi>=-<\Omega_m^{(2)+},\partial_0\psi>=-\int(\partial_0\psi)(q)\,\Omega_m^{(2)+}(dq).
\end{equation}
Let $Y$ be the space of functions $\psi:{\bf R}^4\rightarrow{\bf C}$ such that $\partial_0\psi$ exists for $\Omega_m^{(2)+}$ almost all $q$ in the mass shell $H_m$ and
\[ \int|(\partial_0\psi)(q)|\,\Omega_m^{(2)+}(dq)<\infty. \]
Then $\partial_0\Omega_m^{(2)+}$ can be extended in a natural way to a linear functional $\partial_0\Omega_m^{(2)+}:Y\rightarrow{\bf C}$ defined by the same equation Eq.~\ref{eq:partial_Omega_def} given above defining its action on ${\mathcal S}({\bf R}^4,{\bf C})$.

If $\Upsilon$ is a hyper-rectangle then, using the product rule and the fact that 
\[ (\forall k\in{\bf R}^4)((\partial\chi_{\Upsilon})(q+k)\mbox{ exists and vanishes for $\Omega_m^{(2)+}$ almost all $q\in H_m$}), \]
we can conclude that 
\begin{equation}\label{eq:Xi_1_distribution}
<\Xi_{1,m}^{\mu},\chi_{\Upsilon}>=\frac{1}{2}\pi^2\int\chi_{\Upsilon}(q+k)\frac{\partial}{\partial q^0}(\gamma^{\nu}({\slas q}+m)\gamma^{\mu}({\slas q}-m)\gamma_{\nu})\,\Omega_m^{(2)+}(dq)\,\Omega_0^{+}(dk).
\end{equation}
The RHS of Eq.~\ref{eq:Xi_1_distribution} is easily shown to be a measure as a function of $\Upsilon$ for $\Upsilon$ an arbitrary element of ${\mathcal B}_0({\bf R}^4)$. Therefore we define the measure associated with $\Xi_{1,m}^{\mu}$ to be defined by
\begin{equation}
\Xi_{1,m}^{\mu}(\Upsilon)=\frac{1}{2}\pi^2\int\chi_{\Upsilon}(q+k)\frac{\partial}{\partial q^0}(\gamma^{\nu}({\slas q}+m)\gamma^{\mu}({\slas q}-m)\gamma_{\nu})\,\Omega_m^{(2)+}(dq)\,\Omega_0^{+}(dk).
\end{equation}
It is straightforward to show that $\Xi_m^{\mu}=\Xi_{1,m}^{\mu}+\Xi_{2,m}^{\mu}$ is causal and also, using the Lorentz invariance of $\Omega_m^{(3)+}+\partial_0\Omega_m^{(2)+}$ and $\Omega_0^{+}$ together with the fundamental intertwining property of the Feynman slash \cite{AMP,IJMPA}, that $\Xi_m^{\mu}$ is $K$ covariant.

\subsection{Determination of the density $\Phi_m^{\mu}$}

We now use the spectral calculus to compute the spectrum of the measure $\Xi_{m}^{\mu}$ and then use this spectrum to compute the density for $\Phi_m^{\mu}$.
Using the gamma matrix contraction identities we have
\begin{equation} \label{eq:contracted_expression}
\gamma^{\nu}({\slas q}+m)\gamma^{\mu}({\slas q}-m)\gamma_{\nu}=-2{\slas q}\gamma^{\mu}{\slas q}+2m^2\gamma^{\mu}.
\end{equation}
Thus
\begin{equation} \label{eq:differentiated_expression}
\partial_0(\gamma^{\nu}({\slas q}+m)\gamma^{\mu}({\slas q}-m)\gamma_{\nu})=-2\gamma^0\gamma^{\mu}{\slas q}-2{\slas q}\gamma^{\mu}\gamma^0.
\end{equation}
Hence we have, for $\Xi_{1,m}^{\mu}$,
\begin{align*}
g_{1,m}^{\mu}(a,b,\epsilon)=&\Xi_{1,m}^{\mu}(\Upsilon(a,b,\epsilon))\\
=&\frac{1}{2}\pi^2\int\chi_{\Upsilon(a,b,\epsilon)}(q+k)\frac{\partial}{\partial q^0}(\gamma^{\nu}({\slas q}+m)\gamma^{\mu}({\slas q}-m)\gamma_{\nu})\,\Omega_m^{(2)+}(dq)\,\Omega_0^{+}(dk)\\
\approx&\frac{1}{2}\pi^2\int\chi_{(a,b)}(q^0+k^0)\chi_{B_{\epsilon}({\vct0})}({\vct q}+{\vct k})\frac{\partial}{\partial q^0}(\gamma^{\nu}({\slas q}+m)\gamma^{\mu}({\slas q}-m)\gamma_{\nu})\,\\
&\Omega_m^{(2)+}(dq)\,\Omega_0^{+}(dk)\\
=&-\pi^2\int\chi_{(a,b)}(q^0+k^0)\chi_{B_{\epsilon}({\vct0})}({\vct q}+{\vct k})(\gamma^0\gamma^{\mu}{\slas q}+{\slas q}\gamma^{\mu}\gamma^0)\,\Omega_m^{(2)+}(dq)\,\Omega_0^{+}(dk)\\
=&-\pi^2\int\chi_{(a,b)}(\omega_m({\vct q})+|{\vct k}|)\chi_{B_{\epsilon}({\vct0})}({\vct q}+{\vct k})(\gamma^0\gamma^{\mu}{\slas q}+{\slas q}\gamma^{\mu}\gamma^0)\,\omega_m({\vct q})^{-2}|{\vct k}|^{-1}\,d{\vct q}\,d{\vct k}\\
=&-\pi^2\int\chi_{(a,b)}(\omega_m({\vct q})+|{\vct k}|)\chi_{B_{\epsilon}({\vct0})-{\vct q}}({\vct k})(\gamma^0\gamma^{\mu}{\slas q}+{\slas q}\gamma^{\mu}\gamma^0)\,\omega_m({\vct q})^{-2}|{\vct k}|^{-1}\,d{\vct k}\,d{\vct q}\\
\approx&-\pi^2\int\chi_{(a,b)}(\omega_m({\vct q})+|{\vct q}|)(\gamma^0\gamma^{\mu}{\slas q}+{\slas q}\gamma^{\mu}\gamma^0)\,\omega_m({\vct q})^{-2}|{\vct q}|^{-1}\,d{\vct q}\,(\frac{4}{3}\pi\epsilon^3).
\end{align*}
Therefore
\begin{align*}
g_{1,m,a}^{\mu}(b)=&\lim_{\epsilon\rightarrow0}\epsilon^{-3}g_{1,m}^{\mu}(a,b,\epsilon)\\
=&-\pi^2\int\chi_{(a,b)}(\omega_m({\vct q})+|{\vct q}|)(\gamma^0\gamma^{\mu}{\slas q}+{\slas q}\gamma^{\mu}\gamma^0)(m^2+{\vct q}^2)^{-1}|{\vct q}|^{-1}\,d{\vct q}(\frac{4}{3}\pi).
\end{align*}
Therefore, using spherical polar coordinates, we have that
\begin{align*}
g_{1,m,a}^{\mu}(b)=&-\pi^2\int_{s=Z_m(a)}^{Z_m(b)}\int_{\theta=0}^{\pi}\int_{\phi=0}^{2\pi}(\gamma^0\gamma^{\mu}{\slas q}+{\slas q}\gamma^{\mu}\gamma^0)(m^2+s^2)^{-1}\,s\sin(\theta)\,d\phi\,d\theta\,ds\,(\frac{4}{3}\pi),\\
&\mbox{where }q=(\omega_m(s),{\vct q}),\omega_m(s)=(m^2+s^2)^{\frac{1}{2}},{\vct q}=s(\sin(\theta)\cos(\phi),\sin(\theta)\sin(\phi),\cos(\theta)),
\end{align*}
and
\begin{equation}
Z_m(s)=\frac{s^2-m^2}{2s}.
\end{equation}
Now
\[ \int_{\theta=0}^{\pi}\int_{\phi=0}^{2\pi}{\slas q}\sin(\theta)\,d\phi\,d\theta = 4\pi\omega_m(s)\gamma^0. \]
Thus
\begin{equation}
g_{1,m,a}^{\mu}(b)=-8\pi^3\int_{s=Z_m(a)}^{Z_m(b)}\gamma^0\gamma^{\mu}\gamma^0\omega_m(s)^{-1}s\,ds\,(\frac{4}{3}\pi).
\end{equation}
Let
\begin{equation}
M_{1,m}^{\mu}(b)=\frac{3}{4\pi}g_{1,m,a}^{\mu\prime}(b),
\end{equation}
Then
\begin{equation}\label{eq:M_1}
M_{1,m}^{\mu}(b)=f_{1,m}(b)\gamma^0\gamma^{\mu}\gamma^0=-f_{1,m}(b)\gamma^{\mu}+2f_{1,m}(b)\gamma^0\eta^{0\mu},
\end{equation}
where
\begin{equation}
f_{1,m}(b)=-8\pi^3\left.\frac{}{}(\omega_m(s)^{-1}s)\right|_{s=Z_m(b)}Z_m^{\prime}(b),
\end{equation}
for $b>0$. 

Also we compute using the spectral calculus, for $\Xi_{2,m}^{\mu}$, 
\begin{align*}
g_{2,m}^{\mu}(a,b,\epsilon)=&\Xi_{2,m}^{\mu}(\Upsilon(a,b,\epsilon))\\
=&-\frac{1}{2}\pi^2\int\chi_{\Upsilon(a,b,\epsilon)}(q+k)(-2{\slas q}\gamma^{\mu}{\slas q}+2m^2\gamma^{\mu})\omega_m({\vct q})^{-3}|{\vct k}|^{-1}\,d{\vct k}\,d{\vct q}\\
\approx&\pi^2\int\chi_{(a,b)}(\omega_m({\vct q})+|{\vct k}|)\chi_{B_{\epsilon}({\vct 0})}({\vct q}+{\vct k})({\slas q}\gamma^{\mu}{\slas q}-m^2\gamma^{\mu})\omega_m({\vct q})^{-3}|{\vct k}|^{-1}\,d{\vct k}\,d{\vct q}\\
=&\pi^2\int\chi_{(a,b)}(\omega_m({\vct q})+|{\vct k}|)\chi_{B_{\epsilon}({\vct 0})-{\vct q}}({\vct k})({\slas q}\gamma^{\mu}{\slas q}-m^2\gamma^{\mu})\omega_m({\vct q})^{-3}|{\vct k}|^{-1}\,d{\vct k}\,d{\vct q}\\
=&\pi^2\int\chi_{(a,b)}(\omega_m({\vct q})+|{\vct q}|)({\slas q}\gamma^{\mu}{\slas q}-m^2\gamma^{\mu})\omega_m({\vct q})^{-3}|{\vct q}|^{-1}\,d{\vct q}\,(\frac{4}{3}\pi\epsilon^3)\\
=&\pi^2\int_{s=Z_m(a)}^{Z_m(b)}\int_{\theta=0}^{\pi}\int_{\phi=0}^{2\pi}({\slas q}\gamma^{\mu}{\slas q}-m^2\gamma^{\mu})\omega_m(s)^{-3}s^{-1}s^2\sin(\theta)\,d\phi\,d\theta\,ds\,(\frac{4}{3}\pi\epsilon^3).
\end{align*}
Thus
\begin{equation}
g_{2,m,a}^{\mu}(b)=\pi^2\int_{s=Z_m(a)}^{Z_m(b)}\int_{\theta=0}^{\pi}\int_{\phi=0}^{2\pi}({\slas q}\gamma^{\mu}{\slas q}-m^2\gamma^{\mu})\omega_m(s)^{-3}s\sin(\theta)\,d\phi\,d\theta\,ds\,(\frac{4}{3}\pi).
\end{equation}
From Eq.~\eqref{eq:phi_theta_integral} 
\[ \int_{\theta=0}^{\pi}\int_{\phi=0}^{2\pi}{\slas q}\gamma^{\mu}{\slas q}\sin(\theta)\,d\phi\,d\theta=4\pi(m^2+s^2)\gamma^0\gamma^{\mu}\gamma^0+\frac{4}{3}\pi s^2\gamma^{\mu}+\frac{8}{3}\pi s^2\gamma^0\eta^{\mu0}. \]
Thus
\begin{align*}
g_{2,m,a}^{\mu}(b)=&\pi^2\int_{s=Z_m(a)}^{Z_m(b)}(4\pi(m^2+s^2)\gamma^0\gamma^{\mu}\gamma^0+\frac{4}{3}\pi s^2\gamma^{\mu}+\frac{8}{3}\pi s^2\gamma^0\eta^{\mu0}-4\pi m^2\gamma^{\mu})\\
&\omega_m(s)^{-3}s\,ds\,(\frac{4}{3}\pi).
\end{align*}
Let
\begin{equation}
M_{2,m}^{\mu}(b)=\frac{3}{4\pi}g_{2,m,a}^{\mu\prime}(b).
\end{equation}
Then
\begin{align} \label{eq:M_2}
M_{2,m,a}^{\mu}(b)=&f_{2,1,m}(b)\gamma^{\mu}+f_{2,2,m}(b)\gamma^0\gamma^{\mu}\gamma^0+f_{2,3,m}(b)\gamma^0\eta^{0\mu}\\
=&(f_{2,1,m}(b)-f_{2,2,m}(b))\gamma^{\mu}+(f_{2,3,m}(b)+2f_{2,2,m}(b))\gamma^0\eta^{0\mu},\nonumber
\end{align}
where
\begin{align*}
&f_{2,1,m}(b)=\left.\left\{4\pi^3(\frac{1}{3}s^2-m^2)s\omega_m(s)^{-3}\right\}\right|_{s=Z_m(b)}Z_m^{\prime}(b),\\
&f_{2,2,m}(b)=\left.\left\{\frac{}{}4\pi^3(m^2+s^2)s\omega_m(s)^{-3}\right\}\right|_{s=Z_m(b)}Z_m^{\prime}(b),\\
&f_{2.3,m}(b)=\left.\left\{\frac{8}{3}\pi^3s^2s\omega_m(s)^{-3}\right\}\right|_{s=Z_m(b)}Z_m^{\prime}(b),
\end{align*}
for $b>0$. 

On examination of Eqns.~\eqref{eq:density}, \eqref{eq:spectrum}, \eqref{eq:M_1} and \eqref{eq:M_2} we see that the measure $\Xi_{1,m}$ is associated with the causal $K$ covariant measure $\Psi_{1,m}^{\mu}$ with density
\begin{equation}
\Psi_{1,m}(q)=-f_{1,m}(Q)\gamma^{\mu}+2Q^{-2}f_{1,m}(Q){\slas q}q^{\mu}, 
\end{equation}
and the measure $\Xi_{2,m}^{\mu}$ is associated with a causal $K$ covariant measure $\Psi_{2,m}^{\mu}$ with density
\begin{equation}
\Psi_{2,m}(q)=(f_{2,1,m}(Q)-f_{2,2,m}(Q))\gamma^{\mu}+Q^{-2}(f_{2,3,m}(Q)+2f_{2,2,m}(Q)){\slas q}q^{\mu},
\end{equation}
where $Q=(q^2)^{\frac{1}{2}}$ in such a way that the measure $\Xi_m^{\mu}$, which is causal and $K$ covariant, is the sum
\begin{equation}
\Xi_m^{\mu}=\Psi_{1,m}^{\mu}+\Psi_{2,m}^{\mu}.
\end{equation}
From general principles \cite{Symmetry} if we have a causal Lorentz invariant measure $\mu$ on Minkowski space with spectrum $\sigma:(0,\infty)\rightarrow{\bf C}$ then the spacelike measure associated with $\mu$ has spectrum $\pm\sigma$, the sign being determined by the sign taken for a certain square root, i.e. the branch taken for a certain Riemann surface. We generalize this by saying that if we have a causal covariant measure on Minkowski space with spectral functions $\sigma_i$ then the associated spacelike covariant measure has spectral functions $\pm\sigma_i$.  

Thus, to determine the spacelike measure associated with the measure $\Phi_m^{\mu}=\frac{e^2}{(2\pi)^4}(\Psi_{1,m}^{\mu}+\Psi_{2,m}^{\mu})$ we need to choose four signs in the equations
\begin{align}
\Psi_{1,m}(q)=&\pm f_{1,m}(Q)\gamma^{\mu}\pm2Q^{-2}f_{1,m}(Q){\slas q}q^{\mu}, \\
\Psi_{2,m}(q)=&\pm(f_{2,1,m}(Q)-f_{2,2,m}(Q))\gamma^{\mu}\pm Q^{-2}(f_{2,3,m}(Q)+2f_{2,2,m}(Q)){\slas q}q^{\mu}. \nonumber
\end{align}  
We choose the first two to be negative and positive respectively and the second two to be positive and negative respectively. Therefore $\Phi_m^{\mu}$ is given, on the spacelike region $\{q\in{\bf R}^4:q^2\leq m^2\}$, by
\begin{align*}
\Phi_m^{\mu}(q)=&\frac{e^2}{(2\pi)^4}\Xi_{m}^{\mu}(q)\\
=&\frac{e^2}{(2\pi)^4}((f_{2,1,m}(Q)-f_{1,m}(Q)-f_{2,2,m}(Q))\gamma^{\mu}+\\
&Q^{-2}(-f_{2,3,m}(Q)+2(f_{1,m}(Q)-f_{2,2,m}(Q)){\slas q}q^{\mu}),\label{eq:Phi_timelike_density}
\end{align*}
where $Q=(-q^2)^{\frac{1}{2}}$.

Therefore, for $q\in{\bf R}^4$ for which $q^2\leq-m^2$
\begin{equation} \label{eq:Phi_m_mu_answer}
\Phi_m^{\mu}(q)=\frac{e^2}{(2\pi)^4}(h_1(Q)\gamma^{\mu}+Q^{-2}h_2(Q){\slas q}q^{\mu}), 
\end{equation}
where
\begin{align*}
h_{1,m}(Q)=&f_{2,1,m}(Q)-f_{1,m}(Q)-f_{2,2,m}(Q),\\
h_{2,m}(Q)=&-f_{2,3,m}(Q)+2(f_{1,m}(Q)-f_{2,2,m}(Q)),
\end{align*}
with $Q=(-q^2)^{\frac{1}{2}}$.

\section{Computation of the vertex correction to the cross section for the process $e^{+}e^{-}\rightarrow\mu^{+}\mu^{-}$ in the high energy limit\label{section:electron_positron_annihilation}}

\subsection{Tree level for the process $e^{+}e^{-}\rightarrow\mu^{+}\mu^{-}$}

\begin{figure} 
\centering
\includegraphics[width=6cm]{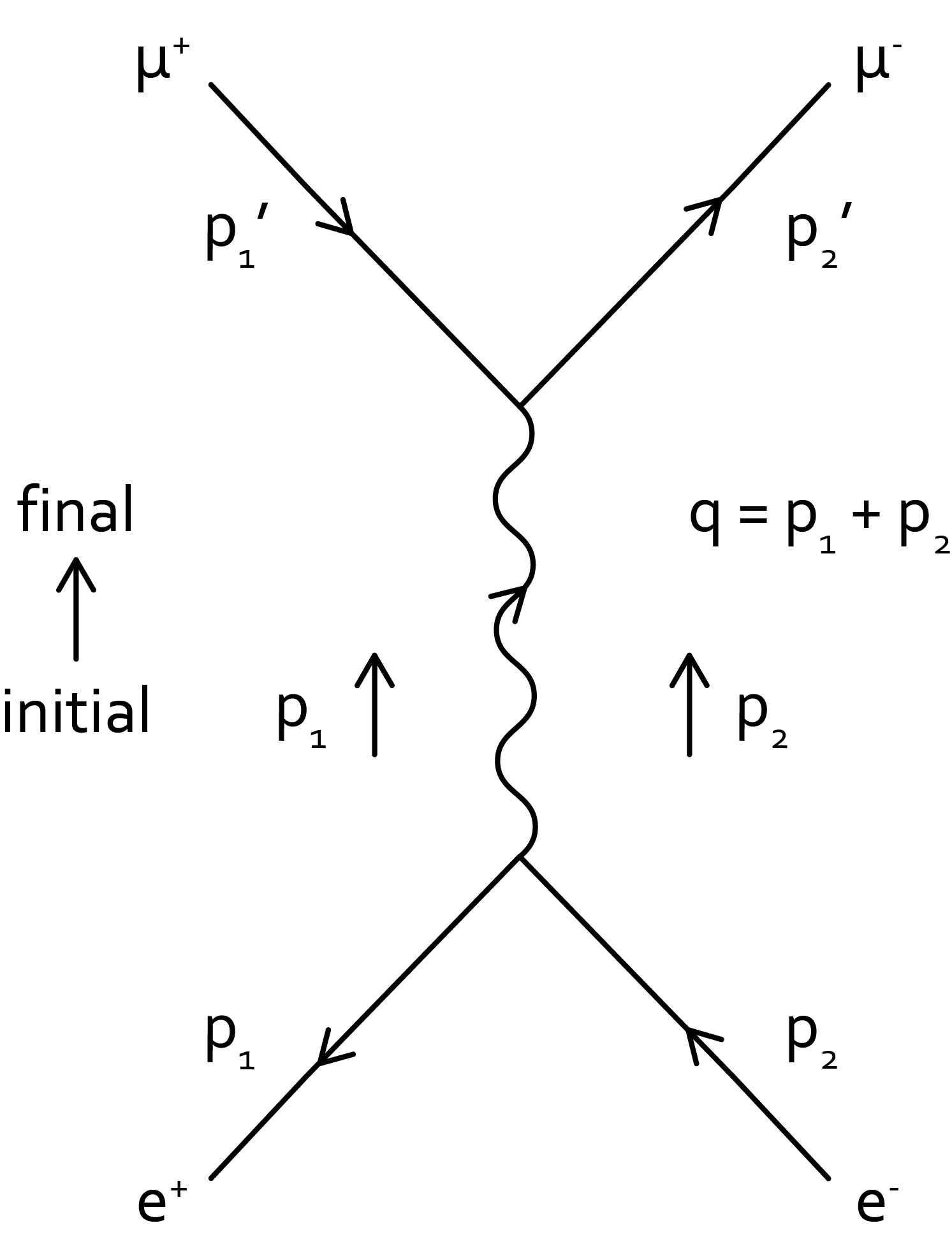}
\caption{Feynman diagram for the process $e^{+}e^{-}\rightarrow\mu^{+}\mu^{-}$ at tree level} \label{figure:s_channel_tree}
\end{figure}

Consider the process $e^{+}e^{-}\rightarrow\mu^{+}\mu^{-}$ whose tree level Feynman diagram is shown in Figure~\ref{figure:s_channel_tree}.  As is usual in such computations the flip $(p_1^{\prime},p_2^{\prime},p_1,p_2)\rightarrow(Tp_1^{\prime},p_2^{\prime},Tp_1,p_2)$ is understood.

Using the Feynman rules the Feynman amplitude ${\mathcal M}_1$ for the tree level diagram is given by
\begin{equation}
i{\mathcal M}_1=\overline{v}_1(p_1,\alpha_1)ie\gamma^{\mu}u_2(p_2,\alpha_2)iD_{\mu\nu}(q)\overline{u}_2^{\prime}(p_2^{\prime},\alpha_2^{\prime})ie\gamma^{\nu}v_1^{\prime}(p_1^{\prime},\alpha_1^{\prime}).
\end{equation}
Therefore
\begin{equation}
{\mathcal M}_1=\frac{e^2}{Q^2}\overline{v}_1(p_1,\alpha_1)\gamma^{\mu}u_2(p_2,\alpha_2)\eta_{\mu\nu}\overline{u}_2^{\prime}(p_2^{\prime},\alpha_2^{\prime})\gamma^{\nu}v_1^{\prime}(p_1^{\prime},\alpha_1^{\prime}),
\end{equation}
where $Q=((p_1+p_2)^2)^{\frac{1}{2}}$.

One can compute \cite{Schwartz} that, in the high energy limit,
\begin{equation}
\overline{|{\mathcal M}_1|^2}=\frac{1}{4}\sum_{\mbox{spins}}{\mathcal M}_1^{\dagger}{\mathcal M}_1=8\frac{e^4}{Q^4}((p_2.p_1^{\prime})(p_2^{\prime}.p_1)+(p_2.p_2^{\prime})(p_1.p_1^{\prime})). \label{eq:tree_answer}
\end{equation}
The associated differential cross section in the CM frame is given by \cite{Schwartz}
\begin{align*}
\left(\frac{d\sigma}{d\Omega}\right)_{CM}=&\frac{1}{64\pi^2E_{CM}^2}\frac{|{\vct p}_2|}{|{\vct p}_1|}\overline{|{\mathcal M}_1|^2}\\
=&\frac{1}{64\pi^2E_{CM}^2}\frac{|{\vct p}_2|}{|{\vct p}_1|}8\frac{e^4}{Q^4}((p_2.p_1^{\prime})(p_2^{\prime}.p_1)+(p_2.p_2^{\prime})(p_1.p_1^{\prime})).
\end{align*}
Now, in this frame, there exists, by a well known theorem, ${\vct p},{\vct p}^{\prime}\in{\bf R}^3$ such that  
\[ p_1=(E,-{\vct p}),p_2=(E,{\vct p}),p_1^{\prime}=(E,-{\vct p}^{\prime}),p_2^{\prime}=(E,{\vct p}^{\prime}),E_{CM}=Q=2E, \].
where $E=\omega_{m_e}({\vct p}_i)=\omega_{m_{\mu}}({\vct p}_i^{\prime}),i=1,2$.
We compute that
\begin{align*}
p_2.p_1^{\prime}=&(E,{\vct p}).(E,-{\vct p}^{\prime})=E^2+{\vct p},{\vct p}^{\prime}\\
p_2^{\prime}.p_1=&(E,{\vct p}^{\prime}).(E,-{\vct p})=E^2+{\vct p}.{\vct p}^{\prime}\\
p_2.p_2^{\prime}=&(E,{\vct p}).(E,{\vct p}^{\prime})=E^2-{\vct p}.{\vct p}^{\prime}\\
p_1.p_1^{\prime}=&(E,-{\vct p}).(E,-{\vct p}^{\prime})=E^2-{\vct p}.{\vct p}^{\prime}.
\end{align*}
Thus
\begin{align*}
(p_2.p_1^{\prime})(p_2^{\prime}.p_1)+(p_2.p_2^{\prime})(p_1.p_1^{\prime}))=&(E^2+({\vct p}.{\vct p}^{\prime}))^2+(E^2-({\vct p}.{\vct p}^{\prime}))^2\\
=&2(E^4+({\vct p}.{\vct p}^{\prime})^2).
\end{align*}
In the high energy limit $E=|{\vct p}_i|=|{\vct p}_i^{\prime}|, i=1,2$ so
\[ (p_2.p_1^{\prime})(p_2^{\prime}.p_1)+(p_2.p_2^{\prime})(p_1.p_1^{\prime}))=2E^4(1+\cos^2(\theta)), \]
where $\theta$ is the angle between ${\vct p}$ and ${\vct p}^{\prime}$and
\begin{align*}
\frac{|{\vct p}_2|}{|{\vct p}_1|}=\frac{E}{E}=1,
\end{align*}
from which it follows that
\begin{align*}
\left(\frac{d\sigma}{d\Omega}\right)_{CM,\mbox{tree}}=&\frac{1}{64\pi^2Q^2}(8\frac{e^4}{Q^4})(2E^4)(1+\cos^2(\theta))\\
=&\frac{e^4}{64\pi^2Q^2}(1+\cos^2(\theta)).\\
\end{align*}
Therefore the cross section $\sigma_0$ for this process at tree level is
given by
\begin{equation}\label{eq:tree_cross-section}
\sigma_0=\frac{e^4}{64\pi^2Q^2}\int_{\theta=0}^{\pi}\int_{\phi=0}^{2\pi}(1+\cos^2(\theta))\sin(\theta)\,d\phi\,d\theta=\frac{e^4}{12\pi Q^2}.
\end{equation}

\subsection{Vertex correction for the process $e^{+}e^{-}\rightarrow\mu^{+}\mu^{-}$}

\begin{figure} 
\centering
\includegraphics[width=6cm]{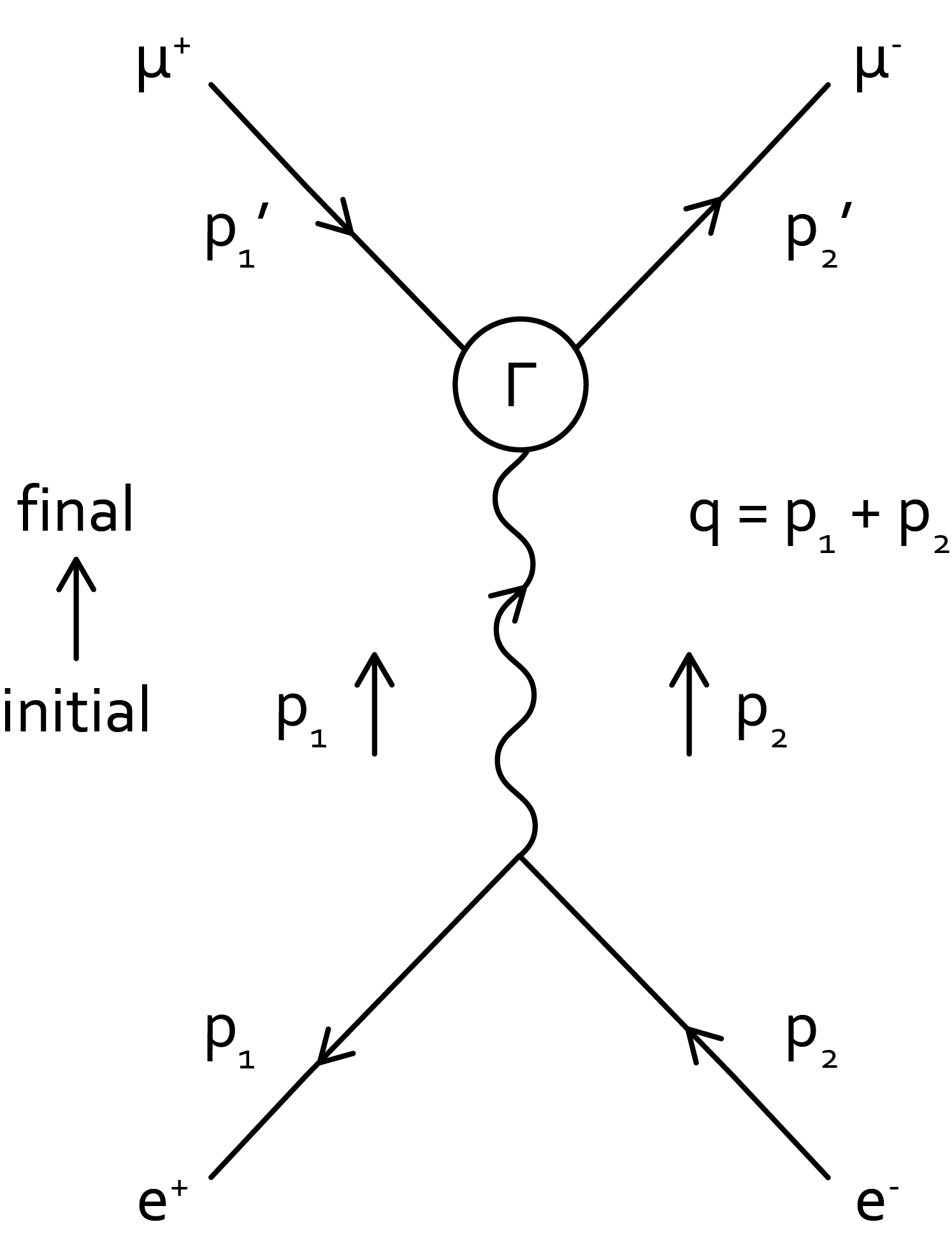}
\caption{Feynman diagram for the process $e^{+}e^{-}\rightarrow\mu^{+}\mu^{-}$ with vertex correction} \label{figure:s_channel_vertex}
\end{figure}

The Feynman diagram for the process $e^{+}e^{-}\rightarrow\mu^{+}\mu^{-}$ with vertex correction is shown in Figure~\ref{figure:s_channel_vertex}.
Using the Feynman rules the Feynman amplitude ${\mathcal M}_2$ associated with this diagram is given by
\begin{equation} \label{eq:M2_def}
i{\mathcal M}_2=\overline{v}_1(p_1,\alpha_1)ie\gamma^{\mu}u_2(p_2,\alpha_2)iD_{\mu\nu}(q)\overline{u}_2^{\prime}(p_2^{\prime},\alpha_2^{\prime})e\Gamma^{\nu}(p_1^{\prime},p_2^{\prime})v_1^{\prime}(p_1^{\prime},\alpha_1^{\prime}).
\end{equation}
If $p_1^{\prime}\in H_{-m}$ and $p_2^{\prime}\in H_m$ are the momenta of an on shell $\mu^{+}$ and an on shell $\mu^{-}$ respectively then $Tp_1^{\prime},p_2^{\prime}\in H_m$. We may define the following  momentum difference/sum 4-vectors
\begin{align*}
q^{(s)}=&p_2^{\prime}-Tp_1^{\prime},\\
q^{(t)}=&p_2^{\prime}+Tp_1^{\prime}.
\end{align*}
$q^{(s)}$ is spacelike and it can be shown that $q^{(s)2}\leq0$ while $q^{(t)}$ is timelike and it can be shown that $q^{(t)2}\geq4m^2$. The $q$ occurring in Eq.~\eqref{eq:M2_def} is the timelike $q=q^{(t)}$.

We define the energy-flip transformation $F:({\bf R}^4)^4\rightarrow({\bf R}^4)^4$ by
\begin{equation}
F(p_1^{\prime},p_2^{\prime},p_1,p_2)=(Tp_1^{\prime},p_2^{\prime},Tp_1,p_2),
\end{equation}
and the momentum-flip operation $G:({\bf R}^4)^4\rightarrow({\bf R}^4)^4$ by
\begin{equation}
G(p_1^{\prime},p_2^{\prime},p_1,p_2)=(Pp_1^{\prime},p_2^{\prime},Pp_1,p_2).
\end{equation}
Let $F(p$ denote the value of a 4-vector $p$ after the energy flip has been carried out. Similarly define $G(p)$ to be the value of a 4-vector $p$ after the momentum flip has been carried out. Under the energy flip transformation the $q$ of Eq.~\eqref{eq:q_def} and $q^{(s)}$ are interchanged. Also we have
\begin{equation}
F(q^{(s)})=p_2^{\prime}-p_1^{\prime}=G(q^{(t)}).
\end{equation}
Thus under an energy-flip the space-like $q^{(s)}$ is interchanged with (G of) the timelike $q^{(t)}$. However we will find that our result is invariant under the operation $G$.
 
The standard calculations in QFT for evaluating the the Feynman amplitude for the process automatically impose an energy-flip $F$ on the momenta.
From this, and also from our calculations in the previous section we can insert $\Phi_{2m}^{\mu}(q)$ with $q=q^{(s)}$ in place of $\Gamma^{\mu}(p_1^{\prime},p_2^{\prime})$ in the  equation \eqref{eq:M2_def} for ${\mathcal M}_2$. However we must multiply $\Phi_{2m}^{\mu}(q)$ by $\frac{i}{\pi}$. The reason for this is that  we are considering two representations of Lorentz invariant measures on Minkowski space. In representation I the measures are synthesized from measures of the form $(p^2-m^2+i\epsilon)^{-1}$. In representation II the measures are synthesized from measures of the form $\Omega_m^{\pm}(p)$. The two representations are related by $(p^2-m^2+i\epsilon)^{-1}\leftrightarrow-i\pi\Omega_m^{\pm}(p)$. i.e. to go from representation I to representation II one must multiply by $-i\pi$. Conversely to go from representation II to representation I one must multiply by $\frac{i}{\pi}$. In our computation of $\Phi_{2m}^{\mu}$ we converted the original equation Eq.~\eqref{eq:original_vertex_equation_s_channel} from representation I to representation II after which we made our computation of the spectrum and density associated with $\Phi_{2m}^{\mu}$. Objects in equations defining Feynman amplitudes must be in representation I. Hence, we must convert the density that we have obtained back to representation I by multiplying it by $\frac{i}{\pi}$. Therefore
\begin{equation}\label{eq:vertex_correction_Feynman_amplitude}
{\mathcal M}_2=\frac{1}{\pi}\frac{e^2}{Q^2}\overline{v}_1(p_1,\alpha_1)\gamma^{\mu}u_2(p_2,\alpha_2)\eta_{\mu\nu}\overline{u}_2^{\prime}(p_2^{\prime},\alpha_2^{\prime})\Phi_{2m}^{\nu}(q)v_1^{\prime}(p_1^{\prime},\alpha_1^{\prime}),
\end{equation}
where $Q=(q^{(t)2})^{\frac{1}{2}}=((Tp_1+p_2)^2)^{\frac{1}{2}}$. 

Let ${\mathcal M}={\mathcal M}_1+{\mathcal M}_2$. Then
\begin{equation}
|{\mathcal M}|^2=|{\mathcal M}_1|^2+|{\mathcal M}_2|^2+{\mathcal M}_1^{\dagger}{\mathcal M}_2+{\mathcal M}_2^{\dagger}{\mathcal M}_1=|{\mathcal M}_1|^2+|{\mathcal M}_2|^2+2\mbox{Re}({\mathcal M}_1^{\dagger}{\mathcal M}_2).
\end{equation}

Using the standard QFT computations involving, for example, the Casimir trick, one can compute that
\begin{align*}
\sum_{\mbox{spins}}{\mathcal M}_2^{\dagger}{\mathcal M}_2=&\frac{1}{\pi^2}\frac{e^4}{Q^4}\eta_{\mu\nu}\eta_{\mu^{\prime}\nu^{\prime}}\mbox{Tr}[({\slas p}_1^{\prime}-m_{\mu})\gamma^0\Phi_{2m}^{\nu^{\prime}}(q)^{\dagger}\gamma^0({\slas p}_2^{\prime}+m_{\mu})\Phi_{2m}^{\nu}(q)]\\
&\mbox{Tr}[({\slas p}_2+m_e)\gamma^{\mu^{\prime}}({\slas p}_1-m_e)\gamma^{\mu}],
\end{align*}
and
\begin{align*}
\sum_{\mbox{spins}}{\mathcal M}_1^{\dagger}{\mathcal M}_2=&\frac{1}{\pi}\frac{e^4}{Q^4}\eta_{\mu\nu}\eta_{\mu^{\prime}\nu^{\prime}}\mbox{Tr}[({\slas p}_1^{\prime}-m_{\mu})\gamma^{\nu^{\prime}}({\slas p}_2^{\prime}+m_{\mu})\Phi_{2m}^{\nu}(q)]\\
&\mbox{Tr}[({\slas p}_2+m_e)\gamma^{\mu^{\prime}}({\slas p}_1-m_e)\gamma^{\mu}],
\end{align*}

Since $\Phi_{2m}^{\mu}$ is is of the order of $e^2$, the ${\mathcal M}_2^{\dagger}{\mathcal M}_2$ contribution to the vertex correction cross section is of the order of $e^8$ ($\alpha^4$) while the ${\mathcal M}_1^{\dagger}{\mathcal M}_2$ contribution is of the order of $e^6$ ($\alpha^3$). Both are finite because we have determined a finite representation for $\Phi_{2m}^{\nu}$. We will compute the second contribution, the LO contribution.

In the high energy limit
\begin{align}
\sum_{\mbox{spins}}{\mathcal M}_1^{\dagger}{\mathcal M}_2=&\frac{e^4}{Q^4}\eta_{\mu\nu}\eta_{\mu^{\prime}\nu^{\prime}}\mbox{Tr}[{\slas p}_1^{\prime}\gamma^{\nu^{\prime}}{\slas p}_2^{\prime}\frac{1}{\pi}\Phi_{2m}^{\nu}(q)]\mbox{Tr}[{\slas p}_2\gamma^{\mu^{\prime}}{\slas p}_1\gamma^{\mu}]\label{eq:leading_order_amplitude}.
\end{align}

We have
\begin{equation}
\frac{1}{\pi}\Phi_{2m}^{\mu}(q)=\frac{1}{\pi}\frac{e^2}{(2\pi)^4}(h_{1,2m}(Q)\gamma^{\mu}+Q^{-2}h_{2,2m}(Q){\slas q}q^{\mu}).
\end{equation}
Let
\begin{equation}
X=\eta_{\mu\nu}\eta_{\mu^{\prime}\nu^{\prime}}YZ,
\end{equation}
where
\begin{align*}
Y=&\mbox{Tr}[{\slas p}_1^{\prime}\gamma^{\nu^{\prime}}{\slas p}_2^{\prime}\frac{1}{\pi}\Phi_{2m}^{\nu}(q)],\\
Z=&\mbox{Tr}[{\slas p}_2\gamma^{\mu^{\prime}}{\slas p}_1\gamma^{\mu}].
\end{align*}
Then 
\[ Y = \frac{1}{\pi}\frac{e^2}{(2\pi)^4}(h_{1,2m}(Q)Y_1+Q^{-2}h_{2,2m}(Q)Y_2), \]
where
\begin{align*}
Y_1=&\mbox{Tr}[{\slas p}_1^{\prime}\gamma^{\nu^{\prime}}{\slas p}_2^{\prime}\gamma^{\nu}],\\
Y_2=&\mbox{Tr}[{\slas p}_1^{\prime}\gamma^{\nu^{\prime}}{\slas p}_2^{\prime}{\slas q}q^{\nu}].
\end{align*}
Now
\begin{align*}
Z=&p_{2\alpha}p_{1\beta}\mbox{Tr}[\gamma^{\alpha}\gamma^{\mu^{\prime}}\gamma^{\beta}\gamma^{\mu}]\\
=&4p_{2\alpha}p_{1\beta}(\eta^{\alpha\mu^{\prime}}\eta^{\beta\mu}-\eta^{\alpha\beta}\eta^{\mu\mu^{\prime}}+\eta^{\alpha\mu}\eta^{\beta\mu^{\prime}})\\
=&4(p_2^{\mu^{\prime}}p_1^{\mu}-(p_1.p_2)\eta^{\mu\mu^{\prime}}+p_2^{\mu}p_1^{\mu^{\prime}}).
\end{align*}
Hence
\[ \eta_{\mu\nu}\eta_{\mu^{\prime}\nu^{\prime}}Z= 4(p_{2\nu^{\prime}}p_{1\nu}-(p_1.p_2)\eta_{\nu\nu^{\prime}}+p_{2\nu}p_{1\nu^{\prime}}). \]
Also
\begin{align*}
Y_1=&\mbox{Tr}[{\slas p}_1^{\prime}\gamma^{\nu^{\prime}}{\slas p}_2^{\prime}\gamma^{\nu}]\\
=&p_{1\alpha}^{\prime}p_{2\beta}^{\prime}\mbox{Tr}[\gamma^{\alpha}\gamma^{\nu^{\prime}}\gamma^{\beta}\gamma^{\nu}]\\
=&4p_{1\alpha}^{\prime}p_{2\beta}^{\prime}(\eta^{\alpha\nu^{\prime}}\eta^{\beta\nu}-\eta^{\alpha\beta}\eta^{\nu\nu^{\prime}}+\eta^{\alpha\nu}\eta^{\beta\nu^{\prime}})\\
=&4(p_1^{\prime\nu^{\prime}}p_2^{\prime\nu}-(p_1^{\prime}.p_2^{\prime})\eta^{\nu\nu^{\prime}}+p_1^{\prime\nu}p_2^{\prime\nu^{\prime}}),\\
&\\
Y_2=&\mbox{Tr}[{\slas p}_1^{\prime}\gamma^{\nu^{\prime}}{\slas p}_2^{\prime}{\slas q}q^{\nu}]\\
=&q^{\nu}(\mbox{Tr}[{\slas p}_1^{\prime}\gamma^{\nu^{\prime}}{\slas p}_2^{\prime}{\slas p}_1^{\prime}]+\mbox{Tr}[{\slas p}_1^{\prime}\gamma^{\nu^{\prime}}{\slas p}_2^{\prime}{\slas p}_2^{\prime}])\\
=&q^{\nu}(p_1^{\prime2}\mbox{Tr}[\gamma^{\nu^{\prime}}{\slas p}_2^{\prime}]+p_2^{\prime2}\mbox{Tr}[{\slas p}_1^{\prime}\gamma^{\nu^{\prime}}])\\
=&4q^{\nu}(p_1^{\prime2}p_2^{\prime\nu^{\prime}}+p_2^{\prime2}p_1^{\prime\nu^{\prime}})\\
=&4(p_1^{\prime\nu}+p_2^{\prime\nu})(p_1^{\prime2}p_2^{\prime\nu^{\prime}}+p_2^{\prime2}p_1^{\prime\nu^{\prime}})\\
=&4(p_1^{\prime2}p_1^{\prime\nu}p_2^{\prime\nu^{\prime}}+p_2^{\prime2}p_1^{\prime\nu}p_1^{\prime\nu^{\prime}}+p_1^{\prime2}p_2^{\prime\nu}p_2^{\prime\nu^{\prime}}+p_2^{\prime2}p_2^{\prime\nu}p_1^{\prime\nu^{\prime}}).
\end{align*}
Let
\[ W_i=\eta_{\mu\nu}\eta_{\mu^{\prime}\nu^{\prime}}ZY_i,i=1,2. \]
Then
\begin{align*}
W_1=& 4(p_{2\nu^{\prime}}p_{1\nu}-(p_1.p_2)\eta_{\nu\nu^{\prime}}+p_{2\nu}p_{1\nu^{\prime}})\\
&4(p_1^{\prime\nu^{\prime}}p_2^{\prime\nu}-(p_1^{\prime}.p_2^{\prime})\eta^{\nu\nu^{\prime}}+p_1^{\prime\nu}p_2^{\prime\nu^{\prime}})\\
=&16[(p_2.p_1^{\prime})(p_1.p_2^{\prime})-(p_1^{\prime}.p_2^{\prime})(p_1.p_2)+(p_2.p_2^{\prime})(p_1.p_1^{\prime})-(p_1.p_2)((p_1^{\prime}.p_2^{\prime})-\\
&4(p_1^{\prime}.p_2^{\prime})+(p_1^{\prime}.p_2^{\prime}))+(p_2.p_2^{\prime})(p_1.p_1^{\prime})-(p_1^{\prime}.p_2^{\prime})(p_2.p_1)+(p_2.p_1^{\prime})(p_1.p_2^{\prime})]\\
=&32((p_2.p_1^{\prime})(p_1.p_2^{\prime})+(p_2.p_2^{\prime})(p_1.p_1^{\prime})),
\end{align*}
and
\begin{align*}
W_2=&4(p_{2\nu^{\prime}}p_{1\nu}-(p_1.p_2)\eta_{\nu\nu^{\prime}}+p_{2\nu}p_{1\nu^{\prime}})\\
&4(p_1^{\prime2}p_1^{\prime\nu}p_2^{\prime\nu^{\prime}}+p_2^{\prime2}p_1^{\prime\nu}p_1^{\prime\nu^{\prime}}+p_1^{\prime2}p_2^{\prime\nu}p_2^{\prime\nu^{\prime}}+p_2^{\prime2}p_2^{\prime\nu}p_1^{\prime\nu^{\prime}})\\
=&16(p_1^{\prime2}(p_2.p_2^{\prime})(p_1.p_1^{\prime})+p_2^{\prime2}(p_2.p_1^{\prime})(p_1.p_1^{\prime})+p_1^{\prime2}(p_2.p_2^{\prime})(p_1.p_1^{\prime})+p_2^{\prime2}(p_2.p_1^{\prime})(p_1.p_2^{\prime})+\\
&(p_1.p_2)(p_1^{\prime2}(p_1^{\prime}.p_2^{\prime})+p_1^{\prime2}p_2^{\prime2}+p_2^{\prime2}p_1^{\prime2}+p_2^{\prime2}(p_1^{\prime}.p_2^{\prime}))+p_1^{\prime2}(p_2.p_1^{\prime})(p_1.p_2^{\prime})+\\
&p_2^{\prime2}(p_2.p_1^{\prime})(p_1.p_1^{\prime})+p_1^{\prime2}(p_2.p_2^{\prime})(p_1.p_2^{\prime})+p_2^{\prime2}(p_2.p_1^{\prime})(p_1.p_2^{\prime}))\\
=&16(3p_1^{\prime2}(p_2.p_2^{\prime})(p_1.p_1^{\prime})+2p_2^{\prime2}(p_2.p_1^{\prime})(p_1.p_1^{\prime})+2p_2^{\prime2}(p_2.p_1^{\prime})(p_1.p_2^{\prime})+p_1^{\prime2}(p_2.p_1^{\prime})(p_1.p_2^{\prime})-\\
&(p_1.p_2)((p_1^{\prime}.p_2^{\prime})(p_1^{\prime2}+p_2^{\prime2})+2p_1^{\prime2}p_2^{\prime2}).
\end{align*}
These calculations have been carried out with respect to the energy-flipped momenta. The simplest way to compute these expressions is to unflip the momenta. Since the measure $\delta(p_1^{\prime}+p_2^{\prime}-p_1-p_2)$ is invariant under Lorentz transformations,
\[ p_1+p_2=p_1^{\prime}+p_2^{\prime}=0,\] 
and so there exists $r\in{\bf R}^4$ such that
\begin{equation}
p_2=-p_1=\frac{1}{2}r.
\end{equation}
Therefore
\begin{align*}
W_1=&32(((\frac{1}{2}r).(-\frac{1}{2}q))((-\frac{1}{2}r).(\frac{1}{2}q))+((\frac{1}{2}r).(\frac{1}{2}q))((-\frac{1}{2}r).(-\frac{1}{2}q)))\\
=&32(\frac{1}{2})^4((-r.q)(-r.q)+(r.q)(r.q))\\
=&4(r.q)^2,
\end{align*}
and
\begin{align*}
W_2=&16(3((-\frac{1}{2}q)^2((\frac{1}{2}r).(\frac{1}{2}q))((-\frac{1}{2}r).(-\frac{1}{2}q))+2(\frac{1}{2}q)^2((\frac{1}{2}r).(-\frac{1}{2}q))((-\frac{1}{2}r).(-\frac{1}{2}q))+\\
&2(\frac{1}{2}q)^2((\frac{1}{2}r).(-\frac{1}{2}q))((-\frac{1}{2}r).(\frac{1}{2}q))+(-\frac{1}{2}q)^2((\frac{1}{2}r).(-\frac{1}{2}q))((-\frac{1}{2}r).(\frac{1}{2}q))-\\
&((-\frac{1}{2}r).(\frac{1}{2}r)((-\frac{1}{2}q).(\frac{1}{2}q))((-\frac{1}{2}q)^2+(\frac{1}{2}q)^2)+2(-\frac{1}{2}q)^2(\frac{1}{2}q)^2\\
=&16(\frac{1}{2})^6(3q^2(r.q)^2-2q^2(r.q)^2+2q^2(r.q)^2+q^2(r.q)^2-r^2((-q^2)(2q^2)+2q^2q^2))\\
=&q^2(r.q)^2.
\end{align*}
To evaluate these results we energy-flip the momenta again. Therefore $q$ becomes $q^{(s)}=p_2^{\prime}-Tp_1^{\prime}$. Thus, in the CM frame 
\begin{equation}
q=(0,2{\vct p}^{\prime}).
\end{equation}
Since $p_1^{\prime}+p_2^{\prime}=p_1+p_2$ and the CM frame of $p_1$ and $p_2$ is the same as the CM frame of $p_1^{\prime}$ and $p_2^{\prime}$ we have by a similar argument, that
\begin{equation}
r=(0,2{\vct p}).
\end{equation} 
Hence $q^2=-4{\vct p}^{\prime2}$ which, since, in the high energy limit $|{\vct p}^{\prime}|=|{\vct p}|=E$, is equal to $-(2E)^2=-Q^2$. Also
\begin{equation}
r.q=-4{\vct p}.{\vct p}^{\prime}=-4|{\vct p}||{\vct p}^{\prime}|\cos(\theta)=-4E^2\cos(\theta),
\end{equation}
where $\theta$ is the angle between ${\vct p}$ and ${\vct p}^{\prime}$.
Thus
\begin{equation}
X=\frac{1}{\pi}\frac{e^2}{(2\pi)^4}(X_1+X_2),
\end{equation}
where
\begin{align*}
X_1=&h_{1,2m}(Q)W_1=4h_{1,2m}(Q)(r.q)^2,\\
X_2=&Q^{-2}h_{2,2m}(Q)W_2=-h_{2,2m}(Q)(r.q)^2.
\end{align*}
Thus
\begin{align*}
X_1+X_2=&(4h_{1,2m}(Q)-h_{2,2m}(Q))(16E^4)\cos^2(\theta)\\
=&Q^4(4h_{1,2m}(Q)-h_{2,2m}(Q))\cos^2(\theta).
\end{align*}
Now, using
\[ Z_m^{\prime}(Q)=\frac{Q^2+m^2}{2Q^2}\rightarrow\frac{1}{2}\mbox{ as }Q\rightarrow\infty, \]
we compute, in the high energy limit
\begin{align*}
f_{1,m}(Q)&\rightarrow-4\pi^3\mbox{ as }Q\rightarrow\infty,\\
f_{2,1,m}(Q)&\rightarrow\frac{2}{3}\pi^3\mbox{ as }Q\rightarrow\infty,\\
f_{2,2,m}(Q)&\rightarrow2\pi^3\mbox{ as }Q\rightarrow\infty,\\
f_{2,3,m}(Q)&\rightarrow\frac{4}{3}\pi^3\mbox{ as }Q\rightarrow\infty,
\end{align*}
(note that the limits are achieved independently of the value of $m$). Therefore
\begin{align*}
&h_{1,2m}(Q)=f_{2,1,2m}(Q)-f_{1,2m}(Q)-f_{2,2,2m}(Q)\rightarrow\frac{2}{3}\pi^3+4\pi^3-2\pi^3=\frac{8}{3}\pi^3\mbox{ as }Q\rightarrow\infty,\\
&h_{2,2m}(Q)=-f_{2,3,2m}(Q)+2(f_{1,2m}(Q)-f_{2,2,2m}(Q))\rightarrow-\frac{4}{3}\pi^3+2(-4\pi^3-2\pi^3)=-\frac{40}{3}\pi^3\mbox{ as }Q\rightarrow\infty.
\end{align*}
Thus
\[ 4h_{1,2m}(Q)-h_{2,2m}(Q)\rightarrow24\pi^3\mbox{ as }Q\rightarrow\infty. \]
Hence
\begin{align*}
\overline{{\mathcal M}_1^{\dagger}{\mathcal M}_2}=&\frac{1}{4}\sum_{\mbox{spins}}{\mathcal M}_1^{\dagger}{\mathcal M}_2\\
=&\frac{1}{4}\frac{e^4}{Q^4}X\\
=&\frac{1}{4}\frac{e^4}{Q^4}\frac{1}{\pi}\frac{e^2}{(2\pi)^4})Q^4(24\pi^3)\cos^2(\theta)\\
=&\frac{3e^6}{8\pi^2}\cos^2(\theta).
\end{align*}
Therefore the LO contribution of the vertex correction to the differential cross section in the high energy limit and the CM frame is given by 
\begin{align*}
\left(\frac{d\sigma}{d\Omega}\right)_{CM,\mbox{vertex}}(Q,\theta,\phi)=&\frac{1}{64\pi^2E_{CM}^2}\frac{|{\vct p}_2|}{|{\vct p}_1|}\overline{{\mathcal M}_1^{\dagger}{\mathcal M}_2+{\mathcal M}_2^{\dagger}{\mathcal M}_1}\\
=&\frac{1}{64\pi^2Q^2}(2\mbox{Re}\overline{({\mathcal M}_1^{\dagger}{\mathcal M}_2)})\\
=&\frac{3e^6}{256\pi^4Q^2}\cos^2(\theta).
\end{align*}
Therefore the LO differential cross section for the process is given by
\begin{align*}
\left(\frac{d\sigma}{d\Omega}\right)_{CM}=&\left(\frac{d\sigma}{d\Omega}\right)_{CM,\mbox{tree}}+\left(\frac{d\sigma}{d\Omega}\right)_{CM,\mbox{vertex}}\\
=&\frac{e^4}{64\pi^2Q^2}(1+\cos^2(\theta))+\frac{3e^6}{256\pi^4Q^2}\cos^2(\theta).
\end{align*}

The contribution of the spectral vertex correction to the total cross section $\sigma=\sigma(Q)$ is given by
\begin{align*}
\sigma=&\int_{\theta=0}^{\pi}\int_{\phi=0}^{2\pi}\left(\frac{d\sigma}{d\Omega}\right)_{CM,\mbox{vertex}}(Q,\theta,\phi)\sin(\theta)\,d\phi\,d\theta\\
=&\frac{3e^6}{256\pi^4Q^2}\int_{\theta=0}^{\pi}\int_{\phi=0}^{2\pi}\cos^2(\theta)\sin(\theta)\,d\phi\,d\theta\\
=&\frac{e^6}{64\pi^3 Q^2}.
\end{align*}
Hence the LO contribution to the total cross section for the process $e^{+}e^{-}\rightarrow\mu^{+}\mu^{-}$ taking into account the vertex correction is
\begin{equation}
\sigma_{\mbox{\small{tot}}}=\frac{e^4}{12\pi Q^2}+\frac{e^6}{64\pi^3Q^2}=\frac{e^4}{12\pi Q^2}(1+\frac{3e^2}{16\pi^2})=\sigma_0(1+\frac{3e^2}{16\pi^2}),
\end{equation}
and we have reproduced the textbook result \cite{Schwartz} for the LO contribution to the $e^{+}e^{-}\rightarrow\mu^{+}\mu^{-}$ total cross section with vertex correction included.

It is to be noted that we have done so without encountering any form of divergence, neither UV nor IR and we have not had to include soft photon final state radiation in order to cancel IR divergence, since there is no IR divergence.

We show in Ref. \cite{Mashford_final_state_radiation}  that, when analyzed using a careful treatment of distributional objects, the final state radiation process $e^{+}e^{-}\rightarrow\mu^{+}\mu^{-}\gamma$ is not associated with any divergence and that, in the soft photon high energy limit, the cross section for the  process vanishes.

\section{Comparison of the spectral differential cross section with the standard (or classical) differential cross section for the process $e^{+}e^{-}\rightarrow\mu^{+}\mu^{-}$ in the high energy limit\label{section:differential_cross_section_comparison}}

\begin{figure} 
\centering
\includegraphics[width=15cm]{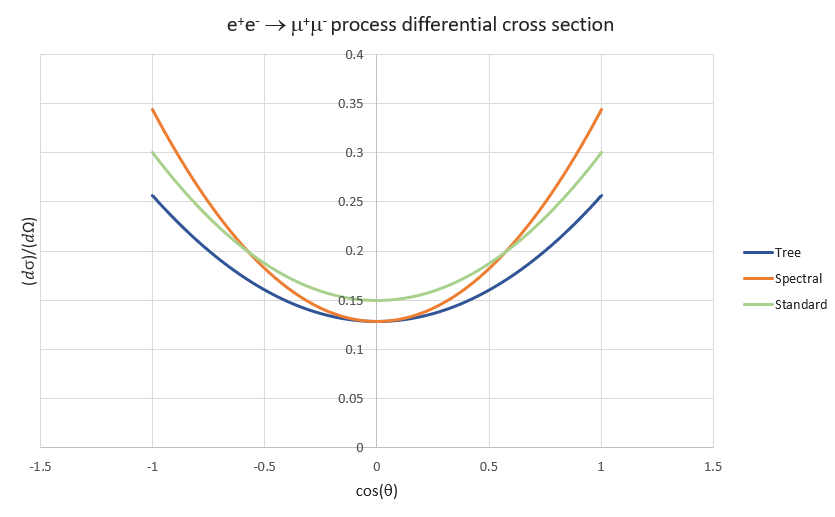}
\caption{Differential cross section for the process $e^{+}e^{-}\rightarrow\mu^{+}\mu^{-}$, both tree and with vertex correction} \label{figure:differential}
\end{figure}

While we hve obtained an identical result for the total cross section, computed using spectral regularization, to the classical result our result for the differential cross section is not the same as the classical result. Figure~\ref{figure:differential} gives graphs of the tree level, the spectral and the classical differential cross sections (to produce these graphs we set $e=3.0$ since with $e=\sqrt{4\pi\alpha}$ the three graphs are so close together as to be indistinguishable). The classical result simply scales up the tree level differential cross section by a constant factor in such a way as to produce the correct total cross section. On the other hand the spectral differential cross section varies with respect to the tree level differential cross section with the total cross section being correct. In particular when $\cos(\theta)=0$ i.e. $\theta=\frac{\pi}{2}$ the spectral differential cross section is equal to the tree level differential cross section.

\subsection{Summary of argument why differential cross section with vertex correction should equal tree level cross section when $\cos(\theta)=0$}

We will now give a general argument why this should be the case. First one must accept the basically well known formula Eq.~\ref{eq:leading_order_amplitude}
\begin{align*}
\sum_{\mbox{spins}}{\mathcal M}_1^{\dagger}{\mathcal M}_2=&\frac{e^4}{Q^4}\eta_{\mu\nu}\eta_{\mu^{\prime}\nu^{\prime}}\mbox{Tr}[{\slas p}_1^{\prime}\gamma^{\nu^{\prime}}{\slas p}_2^{\prime}\frac{1}{\pi}\Phi_{2m}^{\nu}(q)]\mbox{Tr}[{\slas p}_2\gamma^{\mu^{\prime}}{\slas p}_1\gamma^{\mu}],
\end{align*}
required to obtain the LO vertex correction to the cross section in the high energy limit. Second assume that $\Phi^{\nu}$ has the general form
\begin{align*}
\Phi^{\nu}(q)=a(q)\gamma^{\nu}+b(q){\slas q}q^{\nu},
\end{align*}
(this is proved above and the exact forms of the functions $a$ and $b$ are derived). Third one must believe the above computations which show that both the terms in $\sum_{\mbox{spins}}{\mathcal M}_1^{\dagger}{\mathcal M}_2$ arising from $\gamma^{\nu}$ and ${\slas q}q^{\nu}$ respectively are equal to a function of $q$ times $(r.q)$ where $r$ and $q$ are the incoming and outgoing momentum difference 4-vectors. It follows that, using the representation where $r=(0,{\vct r})$ and $q=(0,{\vct q})$ with ${\vct r}$ the incoming 3-momentum and ${\vct q}$ the outgoing 3-momentum, that when $\cos(\theta)=0$ the LO contribution of the vertex correction to the differential cross section vanishes.

If one accepts this general argument (for which all details are provided in the previous sections of this paper) then one can see tha the spectral differential cross section for the process is more reasonable that the standard differential cross section. It is likely that the difference between the two predictions is too small to be detected experimentally. However the difference may be more significant for certain hadronic processes where the coupling constant is much larger.

\section{Conclusion\label{section:conclusion}}

We have computed, without encountering divergence of any sort, analytic functions defining the densities for the vertex function in the t channel and the s channel. These densities can be used in QFT calculations involving the vertex function. In particular, by computing the value at low energy and low momenta of the t channel density we compute the value for the the anomalous magnetic moment of the electron at one loop level and we use the expression for the density in the s channel to compute the LO vertex correction contribution to the high energy limit of the cross section for the process $e^{+}e^{-}\rightarrow\mu^{+}\mu^{-}$ without needing to carry out renormalization or to take into account final state radiation. While the greatest source of uncertainty in the anomalous magnetic moment of the muon seems to come from hadronic contributions \cite{Hagiwara,Lindner} it may be that our computation of the vertex function in addition to our computation of the vacuum polarization tensor \cite{NPB} may have some relevance in regard to solution of the problem of the muon g -- 2 anomaly.\\

%\nocite{*}
%\bibliography{aipsamp}% Produces the bibliography via BibTeX.

\end{document}